\pdfoutput=1

\documentclass[journal]{IEEEtran}

\ifCLASSINFOpdf

\else

\fi

\usepackage{cite}
\usepackage{lipsum}
\usepackage{tikz}
\usetikzlibrary{shapes,arrows}
\usepackage{standalone}
\usepackage{adjustbox}
\usepackage{amsmath,amssymb,amsfonts, mathtools}  
\usepackage{amsthm,bm}
\usepackage{algorithmic}
\usepackage{graphicx}
\usepackage{textcomp}
\usepackage{xcolor}
\usepackage{float}
\usepackage{tabularx,booktabs}
\usepackage[font=small,labelfont=bf]{caption} % Required for specifying captions to tables and figures
\usepackage[subrefformat=parens]{subcaption}
\usepackage{hyperref}
\usepackage{pdfpages}
\usepackage{multirow,makecell}

\DeclareMathOperator*{\argmax}{\arg\!\max}
\usepackage[roman]{parnotes}
\usepackage{algorithm2e,algorithmic}
\newcolumntype{C}{>{\centering\arraybackslash}X}

\begin{document}

% in *personalized* healthcare?
\title{GLYFE: Review and Benchmark of Personalized Glucose Predictive Models in Type-1 Diabetes}

\author{Maxime~De~Bois,
       Mehdi~Ammi,
        and~Moun\^{i}m~A.~El~Yacoubi
\thanks{M. De Bois is with CNRS-LIMSI and the Universit\'{e} Paris-Saclay, Orsay, France (e-mail: maxime.debois@limsi.fr).}
\thanks{M. Ammi is with Universit\'{e} Paris 8, Saint-Denis, France}
\thanks{M. A. El Yacoubi is with Samovar, CNRS, T{\'e}l{\'e}com SudParis, Institut Polytechnique de Paris, \'{E}vry, France}}

% \markboth{Journal of \LaTeX\ Class Files,~Vol.~14, No.~8, August~2015}
% {Shell \MakeLowercase{\textit{et al.}}: Bare Demo of IEEEtran.cls for IEEE Journals}

\maketitle

\IEEEpeerreviewmaketitle

% \documentclass[journal]{IEEEtran}

% \begin{document}

\begin{abstract}

Due to the sensitive nature of diabetes-related data, preventing them from being shared between studies, progress in the field of glucose prediction is hard to assess. To address this issue, we present GLYFE (GLYcemia Forecasting Evaluation), a benchmark of machine-learning-based glucose-predictive models.

To ensure the reproducibility of the results and the usability of the benchmark in the future, we provide extensive details about the data flow. Two datasets are used, the first comprising 10 \textit{in-silico} adults from the UVA/Padova Type 1 Diabetes Metabolic Simulator (T1DMS) and the second being made of 6 real type-1 diabetic patients coming from the OhioT1DM dataset. The predictive models are personalized to the patient and evaluated on 3 different prediction horizons (30, 60, and 120 minutes) with metrics assessing their accuracy and clinical acceptability.

The results of nine different models coming from the glucose-prediction literature are presented. First, they show that standard autoregressive linear models are outclassed by kernel-based non-linear ones and neural networks. In particular, the support vector regression model stands out, being at the same time one of the most accurate and clinically acceptable model. Finally, the relative performances of the models are the same for both datasets. This shows that, even though data simulated by T1DMS are not fully representative of real-world data, they can be used to assess the forecasting ability of the glucose-predictive models.

Those results serve as a basis of comparison for future studies. In a field where data are hard to obtain, and where the comparison of results from different studies is often irrelevant, GLYFE gives the opportunity of gathering researchers around a standardized common environment.

\end{abstract}


% \begin{IEEEkeywords}

% \end{IEEEkeywords}

% \end{document}
\section{Introduction}

\begin{figure*}
    \centering
    \includegraphics[width=\textwidth]{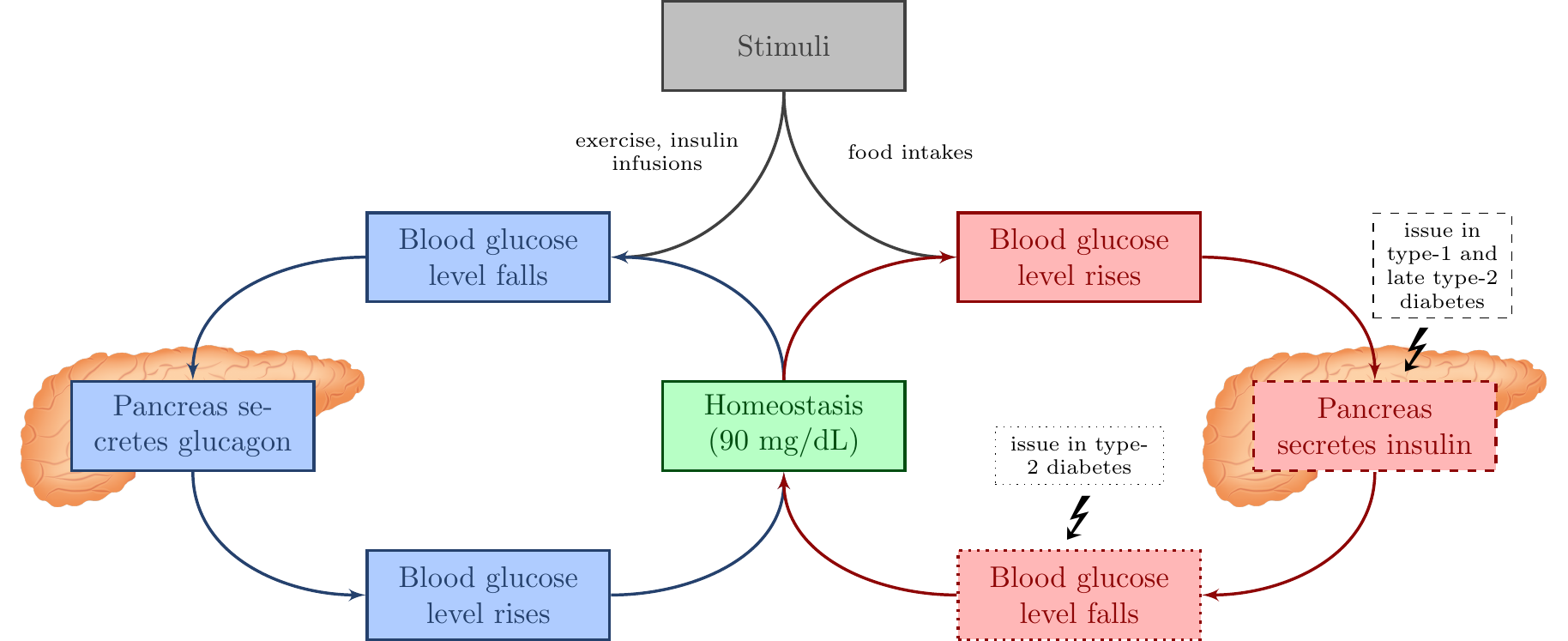}
    \caption{Blood glucose level negative feedback loop.}
    \label{fig:bglregul}
\end{figure*}

Diabetes Mellitus (DM) has become a paramount health issue in the modern world. In 2014, 422 million adults around the world have been estimated to be diabetic and diabetes itself was directly imputed 1.5 million deaths in 2012 \cite{world2016global}.

Diabetes is a chronic disease that can be divided into three main categories: type--1 diabetes mellitus (T1DM), type-2 diabetes mellitus (T2DM) and gestational diabetes. The main problem of diabetic people is the regulation of their blood glucose within acceptable ranges throughout days and nights (homeostasis around 90 $mg/dL$). For a healthy person, it is done automatically by the pancreas that releases insulin and glucagon to counteract the rises and falls of blood glucose level, respectively (see Figure \ref{fig:bglregul}). However, for diabetic people, this negative feedback loop is broken. Whereas in T1DM, the pancreas is unable to produce insulin, in T2DM, the body becomes increasingly resistant to its action (leading eventually to the death of the pancreas' cells responsible for the production of insulin). Consequently, with the help of medication, diabetic people need to manually take care of the regulation of their blood glucose. However, the task is difficult and has severe consequences if not done correctly. Indeed, if the glucose level falls too low (hypoglycemia, below 70 $mg/dL$), it can result in various short-term symptoms including clumsiness, trouble talking, loss of consciousness or even death depending on the severity of the hypoglycemia. On the other hand, high blood glucose (hyperglycemia, above 180 $mg/dL$) can lead to more long-term complications such as poor blood flow, cardiovascular diseases or blindness.

To help diabetic people cope with their disease, a lot of technical solutions have been engineered. One of the most notable and recent progress in the Internet of Things area is the introduction of continuous glucose monitoring (CGM) devices (e.g., FreeStyle Libre \cite{olafsdottir2017clinical}). By using CGM devices, diabetic people do not need to prick their skin with lancets to obtain their current blood glucose level anymore. Besides, we are currently witnessing the rise of smart-phone coaching applications featuring diabetes such as the application mySugr, that has been approved by the Food and Drug Administration (FDA) in the United States \cite{rose2013evaluating}. 

%Mehdi : c'est la partie qui justifie l'interet du projet. je pense qu'il faut développer un peu plus le volet coaching et education therapeutique. et donner des exemple de projets oun de prouduits. il faut peut etre commencer à parler d'horizon temporelle ici pour qu'il comprennent ensuite l'interet de ne pas aller au dela de 1h30. en résumer partie à renforcer.

From a research perspective, the question of predicting future glucose values is of uttermost importance. Making diabetic people know in advance their future glucose levels would help them to anticipate hypo-/hyperglycemia events and thus enable them to take the appropriate action (e.g., eating sugar or taking insulin). Besides, such knowledge could be included into closed-loop systems such as insulin pumps or artificial pancreas to make the insulin delivery more adequate \cite{bequette2012challenges}.

% a lot of advances have been done helping diabetic people regulate their blood glucose. We can identify two main approaches to the problem: building Model Predictive Controllers (MPC), that predict future glucose values and include the prediction in a regulatory decision-based system, and glucose predictive models that predict . In on hand, there is the challenging project of building an Artificial Pancreas (AP) that can automatically deliver the appropriate insulin dosages to the body \cite{bequette2012challenges}. To do so, efforts are focused on building efficient closed-loop systems such as Model Predictive Controllers (MPC) \cite{dassau2012closing}. In the other hand, the endeavors of a lot of researchers have been toward the prediction of future BG values. For instance, the forecasting of future glucose values can then be used inside a MPC or even as is to alert the diabetic of an incoming hypoglycemia.

%Mehdi : je ne pense que soit la bonne tournure pour justifier le simlateur. il faut insister plus sur les interet du simulteur et le fait que ça soit une reference. en gros, les ements sont la, mais la tournure du texte est à revoir je pense. On peut en parler avec mounim notamement.

However, to this day, the question of forecasting glucose levels remains open. The advances in the field are hindered by several factors, one of them being the availability of the data used to train and test the models. First, collecting diabetes-related data, such as the real-time blood glucose level or the insulin boluses taken by the patients, is very time consuming.
Second, those data cannot easily be easily shared among researchers because of their sensitive nature. This leads to the use of datasets that are small and different in studies from one research group to another, making comparisons between them not relevant. Nonetheless, the situation has changed since the development of the UVA/Padova Type 1 Diabetes Metabolic Simulator (T1DMS). Approved by the FDA in the United States as a substitute to pre-clinical animal testing, it simulates the data (glucose levels, insulin boluses, carbohydrate intakes) coming from several \textit{in-silico} patients \cite{man2014uva}. Following its increasing use in glucose prediction research, T1DMS alleviates the burden of data collection. Besides, Marling \textit{et al.} recently released the OhioT1DM dataset made of diverse data such as glucose, insulin and CHO data, but also physical activity, events, and more \cite{marling2018ohiot1dm}. Overall, the glucose prediction research community is presented with a unique oopportunity of building glucose predictive models around the same data. The results of the models can be fully comparable, stimulating the research in the field.

%Mehdi Partie à developper, que aspects et résultats tu veux mettre en évidence, notamement les performances à évaluer, etc.
%c'est le volet motivation et il faut le développer.

%Mehdi 2: je n'arrive pas à trouver facilement la problématique et la question scientifique du papier : la comparaison pour étudier le comprtement des algo en fonction des differentes varaibles du pb ? et ensuite tu dis comment tu traite le pb, donc le benchmark, etc.

How do state-of-the-art glucose predictive models compare with each other in terms of accuracy and clinical acceptability? The goal of this paper is to lay the first stone of results comparison between studies by presenting the benchmark GLYFE (GLYcemia Forecasting Evaluation). Our contributions are:\begin{enumerate}
    \item To initiate the benchmarking process, we implemented nine state-of-the-art glucose predictive models and evaluated them on three different prediction horizons (30, 60, and 120 minutes) and on two different datasets. Results are reported with the Root-Mean-Squared Error (RMSE), that assesses the accuracy of the prediction, the Time-Gain (TG) that estimates the anticipated amount of time gained by forecasting, and the Continuous Glucose-Error Grid Analysis (CG-EGA), that measures the clinical acceptability of the predictions.
    \item To make the results reproducible and to ensure the relevancy of the use of the benchmark in the future, we provide the reader with exhaustive details about the machine-learning pipeline, from the acquisition and preprocessing of the data, to the building of the models and their evaluation.
    \item Compared to former exhaustive reviews that have been done in the field of glucose prediction \cite{oviedo2017review,huzooree2017glucose,woldaregay2019data}, we provide an analysis of the performances of the models. This is made possible by testing the models on a standardized methodology.
    \item The source code of the entire data flow written in Python is provided alongside with detailed benchmark user guidelines through a GitHub repository. It is made flexible so that new models, evaluation metrics, and datasets can easily be added in the future.
    \item Various alternative ways to obtain diabetes-related data exist (such as public datasets or other simulators). We review most of them by showing their strengths and weaknesses.
    % \item  Finally, while several exhaustive reviews have been recently published in the glucose prediction field \cite{oviedo2017review,huzooree2017glucose}, we complete and extend them with some more recent works.

\end{enumerate}

The rest of the paper is structured as follows. First, we review the diabetes public datasets and simulators and glucose predictive models in the literature. Second, we present the whole benchmarking methodology, from the acquisition of the data, to the training, optimization and evaluation of the machine-learning models. Finally, we analyze the benchmark results and conclude.
% ~

\section{State of the Art} \label{sec:state_of_the_art}

\begin{figure*}
    \centering
    \includegraphics[width=\textwidth]{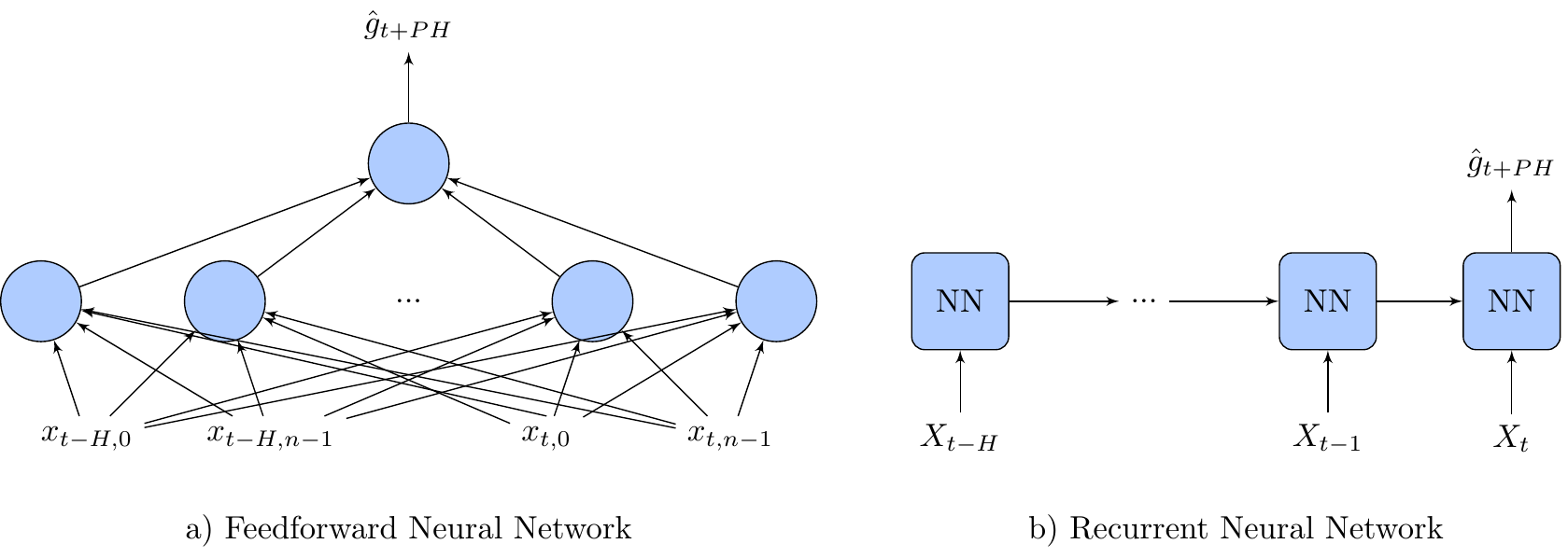}
    \caption{Standard one-layer feedforward neural network (left) and recurrent neural network (right).}
    \label{fig:ffnnrnn}
\end{figure*}

This section reviews the existing public datasets and simulators that can be used in glucose prediction, and covers the state of the art of glucose predictive models.

\subsection{Public Datasets and Simulators}

The most common way of getting data without having to conduct the experiments or data collection is to use public datasets. Besides, recent advances in diabetes metabolic simulation made the emergence of simulators possible \cite{kovatchev2009silico}. This subsection goes through the most popular public datasets and simulators in the field of glucose prediction.

%Mehdi : je pense qu'il faut revoir la strucutre de la première partie de l'etat de l'art. il y a des sections qui portent sur des groupes de recherches et d'autes sur une seule recherche. Je pense qu'il faut qu'une sous sections doit aborder une problematique ou un direction de recherche. par ex. une seciton "simulateur" qui intergre l'ensemble des outils de simulation, une section "datasets" qui regroupe l'ensemble des données existantes, etc. à chaque fois tu développe et montre l'evolution et le fil conducteur. c'est plus le cas de la section "Glucose Predictive Models".

% ~~

\subsubsection{Diabetes Research in Children Network (DirectNet)}

%Mehdi : peut être donner un peu plus de données techniques sur les datasets ?

Since its creation in 2007, DirectNet has been a network whose goal is to investigate the potential use of CGM devices to improve the management of type 1 diabetes in children \cite{ruedy2007diabetes}. To this day, they have conducted multiple studies involving several dozens of subjects (children and adolescents) under various protocols that may include food intake or physical activity. Their work has led to the publication of datasets \cite{directnetstudies} which have been used in the glucose prediction research field \cite{ bazaev2017blood, rudenko2018getting,balakrishnan2014personalized, mhaskar2017deep}. The main downside of these datasets is that they are not representative of the whole type 1 diabetic population because of the instability of glycemic control in diabetic children \cite{jones2003hypoglycemia}. 

% ~~ 

\subsubsection{Diabetes UCI Machine Learning Repository}

The Diabetes Data Set from the UCI Machine Learning Repository \cite{ucidiabetes} has been first introduced for the 1994 AI in Medicine Symposium (AIM-94). It is made of weeks to months of data on 70 diabetic patients. Because of its availability, the dataset has been used outside AIM-94 \cite{khan2017methods,aibinu2010blood,tomczak2016gaussian}. The reason why it has not been used more extensively is because the data are very scarce making the predictions very hard to make. Moreover, the low sampling frequency of glucose data makes the dataset less relevant nowadays. Indeed, through the use of CGM devices, glucose readings can easily be obtained every 5 to 15 minutes, which is not practical through the traditional use of lancets.
% ~~

\subsubsection{AIDA}

AIDA is a software that simulates the effects of insulin and diet changes on blood glucose for T1DM patients \cite{aida}. Started in the early nineties \cite{lehmann1994aida}, the program has been updated throughout the years with its last version published in 2011 (v4.3b). Even though it has been used a few times in the glucose prediction research field \cite{hidalgo2014modeling, reymann2016blood, assadi2017estimation,bamgbose2017closed,mirshekarian2019lstms}, the authors of the simulator claim that, because of the high complexity of the human glucoregulatory system, it is made for educational use only \cite{aida}.

% ~~

\subsubsection{Hovorka Simulation Environment}

Wilinska \textit{et al.} have presented a simulation environment \cite{wilinska2010simulation} based on Hovorka's mathematical model \cite{hovorka2004nonlinear}. By simulating 18 synthetic subjects, the simulator aims at supporting the testing of closed-loop insulin delivery systems. The simulator is mainly used in closed-loop environment to build model predictive controllers \cite{laguna2014experimental,szalay2016long,boiroux2017adaptive}. However, to this day, the simulator is still under clinical trials and has not been fully validated yet.

% ~~

\subsubsection{UVA/Padova T1DMS}

T1DMS, also known as the UVA/Padova Type 1 Diabetes Metabolic Simulator, is a MATLAB-based environment for \textit{in-silico} diabetes modeling. First announced in 2009 \cite{kovatchev2009silico}, it has been updated in 2014 \cite{man2014uva} and is currently undergoing another upgrade \cite{visentin2018uva}. In 2018, it has been accepted by the FDA, in the context of developing new treatments strategies for type-1 diabetes, as a substitute for pre-clinical animal testing. In its public version, T1DMS provides a population of 30 virtual T1DM patients (10 children, 10 adolescents, and 10 adults). During the simulation, every virtual patient is subject to a daily scenario, over which the user has a total control. A scenario can be either open-loop or closed-loop and is defined by meals (timings and quantities), insulin boluses (timings and quantities). In a research field where real-life data are hard to obtain, the simulator has seen a growing use in the recent years \cite{laguna2017enhanced,turksoy2016meal,feng2016performance,li2017blood,contreras2017personalized,contreras2017hybrid,zhao2018multiple,yu2018model,zecchin2012neural,sun2017glucose,sun2018predicting,vehi2019prediction,mirshekarian2019lstms,li2019glunet}. The main downside of using this simulator is that a license must be purchased. Also, there exists an extended version of the simulator with 300 patients which is, to this day, unfortunately not publicly available.

\subsubsection{OhioT1DM Dataset}

In 2018, Marling and Bunescu released a dataset named OhioT1DM for the Blood Glucose Level Prediction (BGLP) Challenge \cite{marling2018ohiot1dm}. The dataset comprises 6 type-1 diabetic patients (identified by theirs IDs: 559, 563, 570, 575, 588, and 591) who have been monitored for 8 weeks in free-living conditions. They wore Medtronic 530G insulin pumps, Medtronic Enlite CGM devices, Basis Peak fitness bands and used a smartphone app to record life events. The following data have been collected: glucose level every 5 minutes from CGM, glucose level from finger sticks, insulin infusions (bolus and basal), meals (times and CHO quantities), exercise, sleep, work, stress, illness, heart rate (every 5 minutes), galvanic skin response, skin temperature, air temperature, and step count. Being the first public dataset to have this variety and per-patient quantity of data, it is seeing a growing interest by the research community \cite{zhu2018deep,bertachi2018prediction,contreras2018using,midroni2018predicting,jeon2019predicting,vehi2019prediction,mirshekarian2019lstms,mayo2019glycemic,li2019glunet,de2019prediction,martinsson2019blood,akbari2019using}

% ~~

\subsection{Glucose Predictive Models}

Extensive reviews of glucose prediction techniques have been done in the past few years \cite{oviedo2017review,huzooree2017glucose,woldaregay2019data}. They showed that we can classify predictive models into three different categories: the physiological models, the data-driven models and the hybrid models. 

% ~~

\subsubsection{Physiological Models}

Physiological models such as the Hovorka model \cite{hovorka2004nonlinear}, the Dalla Man model \cite{dalla2007meal} or the Bergsman minimal model \cite{bergman1979quantitative}, use mathematical equations to describe the human metabolism (food intake, insulin and glucose kinetics).

To forecast the glucose time-series, the Hovorka model is the most used one \cite{laguna2014experimental,calm2011comparison}, followed by the Bergman minimal model \cite{balakrishnan2014personalized,duun2013model}. It should be noted that, to simulate \textit{in-silico} diabetic patients, the Dalla Man model is used in T1DMS and the Hovorka model is used in the Hovorka Simulation Environment.

% ~~

\subsubsection{Data-driven Models}

However, nowadays, researchers have moved away from pure physiological models to time-series analysis and machine learning techniques that show better performances in predicting future glucose values. Various data such as the past glucose values, carbohydrate (CHO) intakes, insulin boluses or measures of physical activity are used to forecast future glucose values. Different prediction horizons (PH) are investigated, with the main ones being 30 (short-term), 60 (mid-term), and 120 minutes (long-term) \cite{oviedo2017review}. Predicting at higher PH is of no value for the patient as the predictions will often be wrong, due to the high number of life events that may occur between the time the prediction is made and the time for which the glucose value is forecast.

Using only the past glucose values, we can predict future ones using an auto-regressive (AR) model \cite{sparacino2007glucose}, which is the most traditionally used glucose predictive model. Equation \ref{eqn:ar} describes the 1-step ahead forecast process using an AR model ($g_i$ and $\hat{g_i}$ are respectively the true and predicted glucose values at time $i$, $p$ is the autoregressive order, and $\boldsymbol{\alpha}$ and $\beta$ the weights of the model). The prediction at a given horizon PH is achieved by predicting one step after another, using the previous prediction as the input to the model. To this model, we can include a moving-average (MA) component to build what we call an ARMA \cite{eren2012adaptive}, or ARIMA \cite{yang2018arima} process. We can incorporate past CHO intakes and insulin boluses into an AR model as exogenous inputs (ARX model) \cite{daskalaki2013early,jankovic2016deep,wang2014personalized,macas2017particle,li2019glunet}. The MA and exogenous components can both be integrated together into an ARMAX process \cite{eren2012adaptive,macas2017particle} or ARIMAX process \cite{novara2016nonlinear}. The ARIMAX includes an Integration (or derivative) component which may be needed when the time-series that is being predicted is non-stationary.

\begin{equation}
\hat{g}_{t+1} = \sum_{i=0}^{p-1} \alpha_i \cdot g_{t-i} + \beta_{t+1}
\label{eqn:ar}    
\end{equation}

After AR models, Artificial Neural Networks (ANN) are the most used models to predict future glucose values. The most common ANN is the Feed-forward Neural Network (FFNN) and has thus been used numerous times to predict future glucose values in diabetes \cite{zecchin2012neural,zarkogianni2015comparative,georga2013glucose,ali2018continuous,vehi2019prediction,mayo2019glycemic}. Recurrent Neural Networks (RNN) are special ANN that are made for time-series forecasting. In RNN, the output (the prediction) of the network at a given time is fed into the inputs of the next prediction, which gives the network a memory of past predictions and therefore, events (see Figure \ref{fig:ffnnrnn}). RNN in its basic form have been widely used in glucose prediction \cite{sandham1998blood,jankovic2016deep,daskalaki2013early}. Besides, some more complex forms of RNN such as RNN with Long Short-Term Memory units (LSTM), extensively used in the machine-learning community, have seen an increased use in recent years \cite{fiorini2017data,mirshekarian2017using, sun2018predicting,de2019prediction,martinsson2019blood,midroni2018predicting,mirshekarian2019lstms}. The idea behind using LSTM units is to cope with the vanishing gradient problem that occurs while training classic RNN \cite{gers1999learning}. Another type of ANN that is quite popular nowadays is the Extreme Learning Machine network (ELM), derived from random vector functional link networks \cite{huang2004extreme, pao1994learning}. With its single hidden layer made of fixed and randomized weighted neurons, it is extremely easy and fast to train. As other ANN, it can be used to build a glucose predictive model \cite{jankovic2016deep,georga2015online,assadi2017estimation,ling2016non}. Other types of ANN have been tried out throughout the years such as neuro-fuzzy neural networks \cite{zarkogianni2015comparative}, self-organizing map \cite{zarkogianni2015comparative}, jump neural networks \cite{zecchin2014jump}, convolutional neural networks \cite{li2019convolutional, zhu2018deep, li2019glunet} or echo state networks \cite{li2018chaotic}.

Another popular approach revolves around the use of kernels to transform the original input space into a higher-dimension space. Known as the \textit{kernel trick}, it enables the learning of non-linear decision boundaries \cite{hofmann2008kernel}. Equation \ref{eqn:rbf} describes the Gaussian kernel (also known as the radial basis function, RBF, kernel), which is one of the most widely used kernel ($x$ and $x'$ being two feature vectors in the original input space, and $\gamma$ a scaling coefficient). In glucose forecasting, this method is used with Gaussian Processes (GP) \cite{tomczak2016gaussian,de2015controlling}, Support Vector Regression (SVR) \cite{bunescu2013blood,georga2013multivariate,georga2013glucose,reymann2016blood,khan2017methods,hamdi2018accurate}, and kernel adaptive filters \cite{naumova2016meta,yu2018online,georga2016non,georga2017kernel,georga2019short}.

\begin{equation}
K(x,x') = \text{exp}(-\gamma\left | \left | x - x' \right | \right |^2)
\label{eqn:rbf}    
\end{equation}

Finally, because of the interpretability of their predictions (which is highly valuable in healthcare), various kinds of decision trees are used (classic, random forests, gradient boosting trees)  \cite{midroni2018predicting,jeon2019predicting,mayo2019glycemic}.

\subsubsection{Hybrid Models}

Hybrid models are models that use compartmental models (taken from physiological models such as Bergman's or Dalla Man's) as input to data-driven models. For instance, Zecchin \textit{et al.} incorporated Dalla Man's compartmental models into an ANN \cite{zecchin2013physical}. The combination of ANN and compartmental models has also been explored by Bertachi \textit{et al.} \cite{bertachi2018prediction}. In their work, Mougiakakou \textit{et al.} presented a hybrid model based on a RNN and Lehmann's equations that describe the insulin and CHO intake kinetics \cite{mougiakakou2006neural}. Finally, Contreras \textit{et al.} developed an approach using grammatical evolution alongside with compartmental models to forecast mid-term future glucose values \cite{contreras2017personalized,contreras2018using}.

\section{GLYFE}

\begin{figure*}
    \centering
    \includegraphics[width=\textwidth]{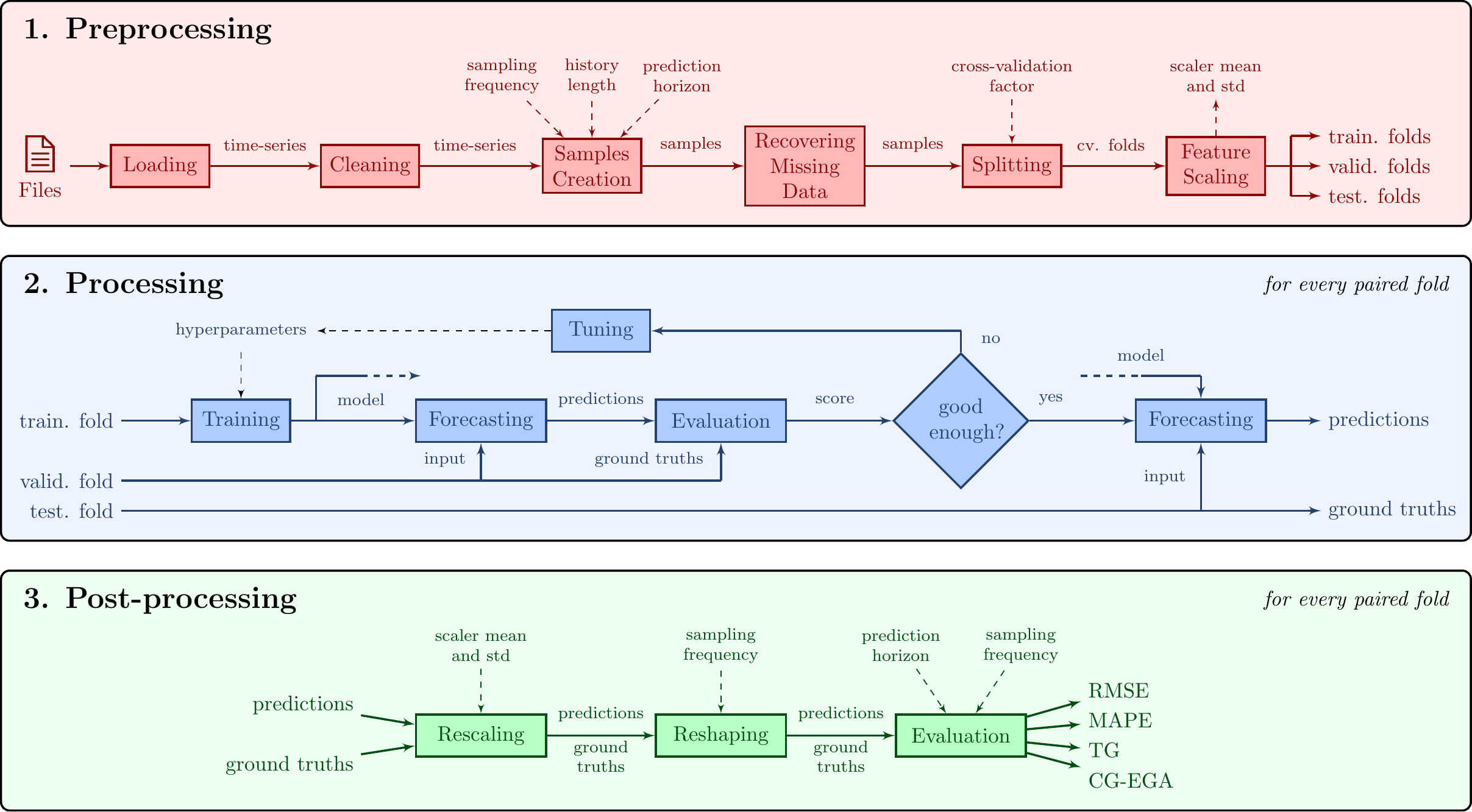}
    \caption{GLYFE benchmark diagram with the preprocessing of the data, the training and tuning of the models, and the evaluation of the predictions.}
    \label{fig:dataflow}
\end{figure*}

This section aims at presenting the whole benchmark methodology. In particular, we go through the data acquisition and preprocessing, the training and tuning of the benchmarked models, and their evaluation. Figure \ref{fig:dataflow} provides a graphical representation of the benchmarking process.

\subsection{Preprocessing}

This section describes the preprocessing and acquisition of the data. The steps are described by the red block \textit{1. Preprocessing} in Figure \ref{fig:dataflow}.

\subsubsection{Loading Data}

Because of their wide research acceptance, the two datasets used in this study come from the T1DMS software and the OhioT1DM dataset. We provide here details on how to obtain them.

The T1DMS data being the property of The Epsilon Group, we must proceed to its whole simulation. It is done on the 10-adult type-1 diabetic population of the software. Every patient, while using generic insulin pumps and CGM sensors, were subject to the following daily scenario \cite{daskalaki2012real, zarkogianni2011insulin, zecchin2012neural}:
\begin{enumerate}
\item[$\bullet$] For every day of the simulation, a patient takes 3 meals whose CHO quantities and timings are randomized. In particular, the timings have been sampled on Gaussian distributions with a variance of 0.5 and means of 7h, 13h and 20h respectively. As for the quantities, they have also been sampled from normal distributions with mean of 40g, 85g and 60g respectively, and variance of 0.5 times the mean CHO quantity of the given meal. Every meal lasts 15 minutes.
\item[$\bullet$] At the start of every meal, an insulin bolus is taken. The bolus value is taken uniformly between 0.7 and 1.3 times the optimal personalized bolus (computed from the patient's personal carbohydrate-to-insulin ratio).
\item[$\bullet$] Every patient is subject to a constant basal insulin injection, which is optimized by the simulator.
\end{enumerate}
The simulation lasts 8 weeks for the sake of uniformity with the OhioT1DM dataset. The randomization of meals timings, CHO quantities and insulin bolus values accounts for the variability of real-life situations (uncontrolled environment). In the end, for every patient, the simulation outputs three different time-series with a sample every minute: glucose readings (mg/dL), CHO intakes (g/min) and insulin boluses (pmol/min).

To obtain the OhioT1DM dataset, one should refer to the instructions in \cite{marling2018ohiot1dm}. For the sake of uniformity with the T1DMS dataset, we restrict ourselves to glucose readings (mg/dL), CHO intakes (g/min), and insulin infusions (in units). 

\subsubsection{Data Cleaning}

Depending on the dataset, data might possess erroneous sensor readings, wrongly reported information by the patients, or formatting errors. For instance, in our previous research, working with another dataset that required the removal of sensor errors, we proposed a methdology to automatically them \cite{de2019prediction}. This step is optionnal and should be done with caution as modifying the data can introduce biases that will eventually negate the evalution. In this study, as we aim to produce baseline results following a standard methodology, we did not use any data cleaning steps.

\subsubsection{Data Samples Creation}

The original sampling frequency of the T1DMS dataset, one sample every minute, is much higher than actual CGM devices. To make the T1DMS dataset more representative of real data, we resample it to one sample every 5 minutes, which is the sampling frequency of the OhioT1DM dataset. During the resampling, we took the mean of glucose values, and the sum of CHO intakes and insulin infusions.

The data samples that are given to the models for their training and evaluation are made of the 3-hour history of glucose values, CHO intakes, and insulin infusions (input) and the glucose reading at the prediction horizon (output). In this study, three prediction horizons are explored: 30 minutes (short-term), 60 minutes (mid-term), and 120 minutes (long-term).

\subsubsection{Training, Validation, and Testing Sets Splitting}

As it is done traditionally in machine learning, both datasets are split into training, validation, and testing sets, each of them having a different purpose. The training set is used to train the models. The validation set is used to evaluate the model during the optimization of its hyperparameters, to ensure that the learnt model and hyperparameters translate well onto unseen data. The testing set is used for the final evaluation of the models.

In this study, the testing sets of both datasets are the last 10 days of data (traditional testing set of the OhioT1DM dataset \cite{marling2018ohiot1dm}). The remaining days are split into training and validation sets following a 80\%/20\% distribution, according to a 5-fold cross-validation evaluation.

\subsubsection{Recovering Missing Data}

Contrary to the T1DMS dataset, the OhioT1DM dataset possesses a lot of missing glucose readings. Some of those values can be recovered by following this strategy applied to every sample:
\begin{enumerate}
    \item linearly interpolate the glucose history when possible (i.e., when the missing value is surrounded by two known glucose readings);
    \item linearly extrapolate the glucose history if the linear interpolation is not possible (usually when the missing glucose reading is the most recent one);
    \item throw away samples when the ground truth is not known to prevent from training on artificial ground truths.
\end{enumerate}
This cleaning strategy ensures that data from the future are not used (as they are not available in real-life situations) and that the models are evaluated on real observations and not artificial ones.

\subsubsection{Feature Scaling}

At the end of the preprocessing stage, the data are standardized (zero mean and unit variance) w.r.t. their training set. This ensures that the models are evaluated on similar data distributions as they have been trained on.

\subsection{Processing}

The training and tuning of all the models are personalized to the patient. We present here the models that have been benchmarked, providing the hyperparameters and their optimization methodology. In Figure \ref{fig:dataflow}, it corresponds to the blue block \textit{2. Processing}.

\subsubsection{Coarse-to-fine Tuning}

Most of the models tested in this study have been previously used in the context of glucose forecasting. Using the same model hyperparameters as in the state of the art would not be fair to the model, as the experimental settings (e.g., data, preprocessing steps) are different. Instead, we need to optimize them individually. Because the models are personalized to the patient, the hyperparameters of the models need also to be personalized to the patient. Because of the high number of models and patients in this study, it is too costly to perform a uniform and detailed grid search for every hyperparameter. Instead, we propose the following coarse-to-fine tuning methodology for the optimization of the hyperparameters: 
\begin{enumerate}
    \item Based on the state of the art, coarse boundaries for each hyperparameter search space are identified. Hyperparameters, for which changes in value yield no change in performance, have been frozen to a default value.
    \item A shallow grid-search is performed on the identified search space. The scale of the search can either be linear or logarithmic, depending on the hyperparameter.
    \item The best coarse value for each hyperparameter from the previous step is refined by a local search.  
\end{enumerate}

\subsubsection{Models}

We present here the nine models that have been implemented for this benchmark: the Base model that provides a baseline comparison, three traditional regression models (Poly, AR, ARX), two more complex non-linear regression models (SVR, GP), and three neural-network-based models (ELM, FFFN, LSTM).

The \textbf{Base} model predicts a glucose value equal to the value at the time it makes the prediction. This model does not require any training nor tuning. It serves as a baseline model for comparison.

The \textbf{Poly} \cite{sparacino2006continuous} model is a polynomial regression model using only the time of the prediction as input to the model. It is not expected to perform very well because of the day-to-day high variability, especially concerning meal times. The order of the model is optimized in the $[10^0, 10^2]$ range.

The \textbf{AR} \cite{sparacino2007glucose} and the \textbf{ARX} \cite{daskalaki2013early,jankovic2016deep,wang2014personalized,macas2017particle} models come both from the family of ARIMAX models. ARIMAX models have 3 different hyperparameters: the endogenous order $p$, the integration order $d$, the MA order $q$. For both AR and ARX models, we have optimized $p$ in the range $[1, 12]$ since it represents the glucose history available to the model,  kept $d$ to 0, and fixed $q$ to 0 since adding any MA component did not seem to make the models any better. ARX models differ from AR models by using additional exogenous inputs (CHO intakes and insulin boluses). The amount of exogenous inputs given to the ARX model matches its order $p$.

\label{sec:svr} The \textbf{SVR} (Support Vector Machine) model has been implemented using the Radial Basis Function (RBF) kernel to transform the input space \cite{bunescu2013blood,georga2013multivariate,reymann2016blood}. We have optimized the coefficient of the kernel in the range $[10^{-4}, 10^{-2}]$. Whereas the loss has been optimized within $[10^{0}, 10^{3}]$, the wideness of the no-penalty-tube has been optimized in the $[10^{-3}, 10^0]$ range.

% ~~

The \textbf{GP} (Gaussian Process) model uses a dot-product kernel with an inhomogeneity  coefficient of $1$ \cite{debois2018study}. Noise has been added to the observations to help the fitting of the model. The amount of noise has been optimized in the $[10^{-3}, 10^2]$ range. 

% ~~

The \textbf{ELM} (Extreme Learning Machines) \cite{georga2015online,assadi2017estimation,ling2016non} model only has two hyperparameters: the number of neurons (logistic activation function) in its single hidden layer, and the L2 penalty applied to the weights, for regularization. Whereas the number of neurons has been optimized in the range $[2000, 20000]$, the L2 penalty has been searched within the $[10^0, 10^3]$ range.

% ~~

The \textbf{FFNN} (Feed-forward Neural Network) \cite{zecchin2012neural,zarkogianni2015comparative,georga2013glucose,ali2018continuous} model is made of 4 hidden layers of respectively 128, 64, 32, and 16 neurons. The activation function of the neurons is the Scaled Exponential Linear Unit (SELU), which is the ELU with an optimized value of $\alpha$ \cite{klambauer2017self}. The model is then fine-tuned by mini-batch ($1500$) using the Adam optimizer \cite{kingma2014adam} with the Mean-Squared Error (MSE) loss function and a learning rate optimized within $[10^{-4},10^{-2}]$. To prevent the model from overfitting the training set, we stopped the training of the model after 100 epochs of no improvement on the validation set (early stopping).

% ~~

The \textbf{LSTM} (Long Short-Term Memory RNN) \cite{fiorini2017data,mirshekarian2017using, sun2018predicting} is made of 2 hidden layers of 256 long short-term memory units. The model is trained with the Adam optimizer and the MSE loss function as well. Mini-batches of size $50$, and a learning rate automatically grid searched within $[10^{-4}, 10^{-3}]$ have been used. As for the amount of regularization that has been used, we applied L2 penalties to the weights ($10^{-4}$) and early stopping (patience of 50 epochs).

The boundaries of the hyperparameters' search spaces have been identified based on the state of the art. Personalized to the patient, the values used for the final evaluation on the testing set are the values minimizing the MSE on the validation set.

\subsection{Post-processing}

\begin{figure*}
    \centering
    \includegraphics[width=\textwidth]{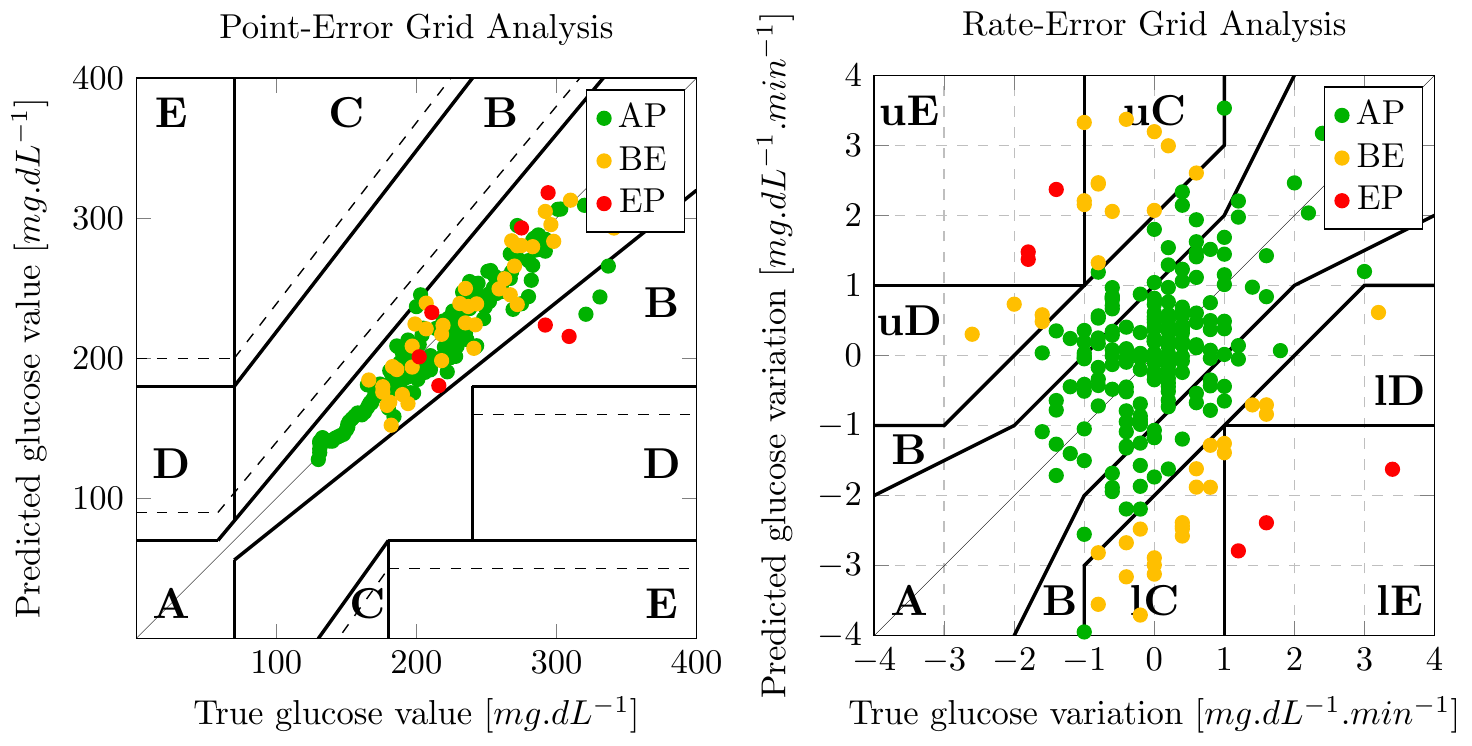}
    \caption{CG-EGA of predictions of the SVR model (see \ref{sec:svr}) for patient 570 from the OhioT1DM dataset for a specific day and a prediction horizon of 30 minutes.}
    \label{fig:cg_ega}
\end{figure*}

The post-processing step is the final step of the benchmarking process (see Figure \ref{fig:dataflow}, represented as the green block \textit{3. Post-processing}). It aims at making the final evaluation of the predictions made by the models.

\subsubsection{Rescaling and Reshaping}

During the preprocessing stage, the glucose values have been scaled to have a zero mean and unit variance. To properly evaluate the predictions, we first need to scale them back to their original mean and variance.

Then, because some of our metrics evaluate consecutive predictions (CG-EGA and TG, see below), we need to reconstruct  the prediction timeline according to the sampling frequency (one prediction every 5 minutes).

\subsubsection{Evaluation Metrics}

Every model is evaluated on the testing set using a 5-fold cross-validation by permutating the training and validation sets. The results are averaged on their respective dataset population for three different prediction horizons: 30 minutes (short-term), 60 minutes (mid-term), and 120 minutes (long-term). Four complementary metrics are used: the Root-Mean-Squared Error (RMSE), the Mean-Absolute-Percentage Error (MAPE), the Time Gain (TG), and the Continuous Glucose-Error Grid Analysis (CG-EGA).

% ~~

Both the RMSE and the MAPE are standard metrics to evaluate the accuracy of time-series forecasting models and are, therefore, widely used in the glucose prediction field \cite{oviedo2017review}. Whereas the RMSE provides a real scale accuracy (see Equation \ref{eqn:rmse}, $\boldsymbol{g}$ and $\boldsymbol{\hat{g}}$ representing respectively the ground truths and the predictions, $N$ the number of samples), the MAPE is scale independent making the comparison between different patients more relevant (see Equation \ref{eqn:mape}).

\begin{equation}
    RMSE(\boldsymbol{g},\boldsymbol{\hat{{g}}})=\sqrt{\frac{1}{N}\cdot\sum_{n=1}^{N} ({g}_{n}-\hat{{g}}_{n})^2}
\label{eqn:rmse}    
\end{equation}

\begin{equation}
    MAPE(\boldsymbol{g},\boldsymbol{\hat{{g}}})=\frac{100}{N} \cdot \sum_{n=1}^{N} \left| \frac{{g}_{n}-\hat{{g}}_{n}}{g_n} \right|
\label{eqn:mape}    
\end{equation}

The TG metric measures the number of minutes gained by the patient through the anticipation of future glucose variations. It is computed as the difference between the prediction horizon and the time-shift (in minutes) that maximizes the correlation between the true and the predicted glucose values (see Equation \ref{eqn:tg}) \cite{gani2009predicting,facchinetti2011new}.

\begin{equation}
TG(\boldsymbol{g},\boldsymbol{\hat{{g}}},PH)=PH-\argmax_{i\in [0,PH]} corr(\boldsymbol{g}_{1...N-i}, \boldsymbol{\hat{g}}_{1+i...N})
\label{eqn:tg}    
\end{equation}

Initially introduced to evaluate the clinical acceptability of CGM devices \cite{kovatchev2004evaluating}, the CG-EGA has seen a lot of use in the glucose prediction research community \cite{oviedo2017review}. The CG-EGA uses two grids evaluating the accuracy of the predictions (P-EGA) and the accuracy of the predicted rates of change (R-EGA). Each grid gives the prediction a rank from A (best) to E (worst) estimating the danger, for the patient, of making such a prediction. Then, depending on the true glycemia region (hypoglycemia, euglycemia\footnote{ Euglycemia is the glycemia region between hypoglycemia and hyperglycemia, or between 70 and 180 mg/dL.}, or hyperglycemia), both marks are used to determine if the prediction is either an Accurate Prediction (AP), a Benign Error (BE), or an Erroneous Prediction (EP). Whereas AP are overall good predictions for the patient, BE are harmless errors, and EP are life-threatening errors. For a model to be clinically acceptable, it needs to have a high AP rate as well as a low EP rate.

Figure \ref{fig:cg_ega} provides a graphical representation of the CG-EGA.

We chose the CG-EGA over other similar metrics (e.g., Clarke Error Grid Analysis \cite{clarke2005original}, Parkes Error Grid \cite{parkes2000new}) as it penalizes the lack of coherence between consecutive predictions, potentially confusing and then dangerous to the patient.

\subsection{Open-Source Software}

 Aiming at making the results reproducible and at promoting its use in the future, we have released the benchmark in a GitHub repository \cite{debois2018GLYFE}.
 
 To use the GLYFE source code, the user first needs to obtain the data. Whereas the OhioT1DM dataset can be accessed through \cite{marling2018ohiot1dm}, the T1DMS data needs to be simulated by the user using the T1DMS software (v3.2.1) in MATLAB (R2018a). To ensure the simulated data are identical to the data used in the benchmark, we provide a step-by-step tutorial that includes: the scenario file, the  parameters of the simulator, the random seed, and the SHA-256 checksum. In particular, the random seed is set twice to 1: first, before launching the simulator by running the \textit{rng(1,"twister")} command in the MATLAB console, and then directly in the simulator interface before running the simulation.
 
 The benchmark has been written in Python 3.7 with the help of standard machine-learning libraries such as Scikit-learn (Poly, SVR, GP, and ELM models) \cite{pedregosa2011scikit}, statsmodels (AR and ARX models) \cite{mckinney2011time}, PyTorch (FFNN and LSTM models) \cite{paszke2017automatic}. 
 
 Following  Figure \ref{fig:dataflow}, the source-code is made of three main modules called \textit{preprocessing}, \textit{processing}, and \textit{post-processing}. The \textit{preprocessing} module contains the general preprocessing functions shown in the figure as well as the dataset-specific functions (e.g., the function that handles the missing data in the OhioT1DM dataset). The \textit{processing} module is made of the general cross-validation training and tuning loop, as well as a folder containing the implementation of the models, and another folder with their respective hyperparameters. Finally, the \textit{post-processing} module contains all the metrics' implementations as well as convenient tools for the visualization of the results.
 
 Overall, the implementation of the whole pipeline described by Figure \ref{fig:dataflow} has been made flexible so that new models, new metrics, new preprocessing steps or even new data can easily be included while preserving,  at the same time, the integrity of the benchmark.

\section{Results \& Discussion}

% \Figure[t!]()[width=\textwidth]{graphs/intra-inter_variability_both.pdf}
% {Per patient (ID) mean and standard deviation of daily MAPE (PH of 30 minutes) for the T1DMS dataset (top) and the OhioT1DM dataset (bottom).\label{fig:intrainter_both}}

Whereas Table \ref{table:acc_t1dms} and Table \ref{table:acc_ohio} present the accuracy-related performances (RMSE, MAPE, TG) of the models for the T1DMS and OhioT1DM datasets respectively, Table \ref{table:cgega_t1dms} and Table \ref{table:cgega_ohio} provide the detailed clinical acceptability of the models (CG-EGA).

\begin{table*}
\parbox{.45\linewidth}{
\caption{Mean accuracy (with standard deviation) of glucose prediction models over the T1DMS population grouped by prediction horizon (30, 60, and 120 minutes).}
\label{table:acc_t1dms}
\begin{tabularx}{\linewidth}{l||C|C|C}
\toprule

\textbf{Models} &  \textbf{RMSE} &\textbf{MAPE} & \textbf{TG}   \\

\midrule
\midrule
\multicolumn{4}{c}{\textbf{\textit{Prediction Horizon = 30 minutes}}}\\
\midrule

\textbf{Ref} & 22.31 \scriptsize{(3.39)} & 12.01 \scriptsize{(1.67)} & 0.00 \scriptsize{(0.00)}\\
\midrule
\textbf{Poly} & 42.94 \scriptsize{(15.40)} & 25.63 \scriptsize{(10.55)} & 24.10 \scriptsize{(2.70)}\\
\textbf{AR} & 13.08 \scriptsize{(1.13)} & 7.77 \scriptsize{(0.83)} & 13.30 \scriptsize{(2.15)}\\
\textbf{ARX} & 11.78 \scriptsize{(0.87)} & 7.21 \scriptsize{(0.78)} & 16.90 \scriptsize{(3.73)}\\
\midrule
\textbf{SVR} & 9.05 \scriptsize{(0.51)} & 5.90 \scriptsize{(0.60)} & 22.70 \scriptsize{(2.37)}\\
\textbf{GP} & \underline{\textbf{9.00 \scriptsize{(0.54)}}} & \underline{\textbf{5.88 \scriptsize{(0.56)}}} & 22.50 \scriptsize{(2.50)}\\
\midrule
\textbf{ELM} & 13.39 \scriptsize{(2.53)} & 7.33 \scriptsize{(0.84)} & \underline{\textbf{24.50 \scriptsize{(1.20)}}}\\
\textbf{FFNN} & 10.37 \scriptsize{(0.73)} & 6.46 \scriptsize{(0.61)} & 19.60 \scriptsize{(2.76)}\\
\textbf{LSTM} & 10.14 \scriptsize{(0.78)} & 6.35 \scriptsize{(0.64)} & 18.60 \scriptsize{(2.11)}\\

\midrule
\midrule
\multicolumn{4}{c}{\textbf{\textit{Horizon de prédiction = 60 minutes}}}\\
\midrule

\textbf{Ref} & 32.63 \scriptsize{(5.78)} & 17.75 \scriptsize{(2.97)} & 0.00 \scriptsize{(0.00)}\\
\midrule
\textbf{Poly} & 42.94 \scriptsize{(15.40)} & 25.63 \scriptsize{(10.55)} & \underline{\textbf{54.10 \scriptsize{(2.70)}}}\\
\textbf{AR} & 24.44 \scriptsize{(3.32)} & 13.94 \scriptsize{(1.98)} & 16.50 \scriptsize{(2.29)}\\
\textbf{ARX} & 22.73 \scriptsize{(2.69)} & 13.38 \scriptsize{(2.02)} & 20.00 \scriptsize{(5.20)}\\

\midrule
\textbf{SVR} & \underline{\textbf{16.07 \scriptsize{(1.53)}}} & \underline{\textbf{9.52 \scriptsize{(1.22)}}} & 44.00 \scriptsize{(2.19)}\\
\textbf{GP} & 16.19 \scriptsize{(1.56)} & 9.84 \scriptsize{(1.25)} & 42.30 \scriptsize{(2.37)}\\
\midrule
\textbf{ELM} & 18.40 \scriptsize{(2.89)} & 10.52 \scriptsize{(1.51)} & 48.00 \scriptsize{(2.14)}\\
\textbf{FFNN} & 17.84 \scriptsize{(1.65)} & 10.82 \scriptsize{(1.29)} & 43.60 \scriptsize{(4.10)}\\
\textbf{LSTM} & 18.66 \scriptsize{(1.95)} & 11.34 \scriptsize{(1.63)} & 37.70 \scriptsize{(4.10)}\\

\midrule
\midrule
\multicolumn{4}{c}{\textbf{\textit{Horizon de prédiction = 120 minutes}}}\\
\midrule

\textbf{Ref} & 46.25 \scriptsize{(9.93)} & 26.53 \scriptsize{(5.75)} & 0.00 \scriptsize{(0.00)}\\
\midrule
\textbf{Poly} & 42.94 \scriptsize{(15.40)} & 25.63 \scriptsize{(10.55)} & \underline{\textbf{114.10 \scriptsize{(2.70)}}}\\
\textbf{AR} & 37.42 \scriptsize{(7.43)} & 22.38 \scriptsize{(4.87)} & 20.10 \scriptsize{(5.94)}\\
\textbf{ARX} & 36.81 \scriptsize{(7.02)} & 22.30 \scriptsize{(4.77)} & 23.90 \scriptsize{(7.54)}\\

\midrule
\textbf{SVR} & 26.55 \scriptsize{(4.96)} & \underline{\textbf{14.26 \scriptsize{(2.56)}}} & 93.80 \scriptsize{(13.20)}\\
\textbf{GP} & 27.27 \scriptsize{(4.74)} & 15.85 \scriptsize{(2.81)} & 72.40 \scriptsize{(10.70)}\\
\midrule
\textbf{ELM} & 26.75 \scriptsize{(4.71)} & 14.81 \scriptsize{(2.59)} & 108.40 \scriptsize{(6.12)}\\
\textbf{FFNN} & \underline{\textbf{25.48 \scriptsize{(4.54)}}} & 14.71 \scriptsize{(2.90)} & 104.80 \scriptsize{(4.79)}\\
\textbf{LSTM} & 32.78 \scriptsize{(5.81)} & 20.21 \scriptsize{(4.14)} & 55.80 \scriptsize{(9.23)}\\

\bottomrule

% \bottomrule
\end{tabularx}
}
\hfill
\parbox{.45\linewidth}{
\centering
\caption{Mean accuracy (with standard deviation) of glucose prediction models over the OhioT1DM population grouped by prediction horizon (30, 60, and 120 minutes).}
\label{table:acc_ohio}
\begin{tabularx}{\linewidth}{l||C|C|C}
\toprule

\textbf{Models} &  \textbf{RMSE} &\textbf{MAPE} & \textbf{TG}   \\

\midrule
\midrule
\multicolumn{4}{c}{\textbf{\textit{Prediction Horizon = 30 minutes}}}\\
\midrule

\textbf{Ref} & 28.32 \scriptsize{(2.38)} & 13.51 \footnotesize{(2.72)} & 0.00 \scriptsize{(0.00)}\\
\midrule
\textbf{Poly} & 57.27 \scriptsize{(6.59)} & 31.09 \scriptsize{(6.71)} & \underline{\textbf{23.17 \scriptsize{(10.01)}}}\\
\textbf{AR} & 20.70 \scriptsize{(2.23)} & 9.62 \scriptsize{(2.26)} & 5.17 \scriptsize{(0.37)}\\
\textbf{ARX} & 20.61 \scriptsize{(2.20)} & 9.59 \scriptsize{(2.19)} & 4.33 \scriptsize{(1.49)}\\
\midrule
\textbf{SVR} & 20.10 \scriptsize{(2.34)} & \underline{\textbf{9.08 \scriptsize{(2.12)}}} & 5.83 \scriptsize{(1.86)}\\
\textbf{GP} & \underline{\textbf{20.01 \scriptsize{(2.33)}}} & 9.16 \scriptsize{(2.16)} & 5.83 \scriptsize{(1.86)}\\
\midrule
\textbf{ELM} & 25.38 \scriptsize{(1.60)} & 11.56 \scriptsize{(2.43)} & 5.67 \scriptsize{(4.15)}\\
\textbf{FFNN} & 21.00 \scriptsize{(2.24)} & 9.33 \scriptsize{(2.19)} & 6.00 \scriptsize{(1.83)}\\
\textbf{LSTM} & 20.46 \scriptsize{(2.08)} & 9.24 \scriptsize{(2.10)} & 6.17 \scriptsize{(1.86)}\\

\midrule
\midrule
\multicolumn{4}{c}{\textbf{\textit{Horizon de prédiction = 60 minutes}}}\\
\midrule

\textbf{Ref} & 41.02 \scriptsize{(2.80)} & 20.37 \scriptsize{(3.87)} & 0.00 \scriptsize{(0.00)}\\
\midrule
\textbf{Poly} & 57.27 \scriptsize{(6.58)} & 31.04 \scriptsize{(6.67)} & \underline{\textbf{51.33 \scriptsize{(13.79)}}}\\
\textbf{AR} & 33.20 \scriptsize{(2.69)} & 16.73 \scriptsize{(3.94)} & 6.33 \scriptsize{(2.98)}\\
\textbf{ARX} & 33.43 \scriptsize{(2.53)} & 16.73 \scriptsize{(3.95)} & 5.83 \scriptsize{(1.86)}\\

\midrule
\textbf{SVR} & 32.27 \scriptsize{(2.35)} & \underline{\textbf{15.38 \scriptsize{(3.43)}}} & 10.17 \scriptsize{(6.99)}\\
\textbf{GP} & \underline{\textbf{31.97 \scriptsize{(2.55)}}} & 15.92 \scriptsize{(3.80)} & 9.33 \scriptsize{(5.12)}\\
\midrule
\textbf{ELM} & 35.14 \scriptsize{(2.15)} & 16.91 \scriptsize{(3.69)} & 12.67 \scriptsize{(9.21)}\\
\textbf{FFNN} & 32.93 \scriptsize{(2.58)} & 15.83 \scriptsize{(3.97)} & 10.00 \scriptsize{(3.70)}\\
\textbf{LSTM} & 32.88 \scriptsize{(2.66)} & 16.00 \scriptsize{(3.57)} & 9.67 \scriptsize{(3.94)}\\

\midrule
\midrule
\multicolumn{4}{c}{\textbf{\textit{Horizon de prédiction = 120 minutes}}}\\
\midrule

\textbf{Ref} & 57.65 \scriptsize{(3.57)} & 30.16 \scriptsize{(4.94)} & 0.00 \scriptsize{(0.00)}\\
\midrule
\textbf{Poly} & 57.26 \scriptsize{(6.59)} & 31.01 \scriptsize{(6.64)} & \underline{\textbf{109.33 \scriptsize{(18.45)}}}\\
\textbf{AR} & 47.48 \scriptsize{(3.75)} & 25.69 \scriptsize{(5.79)} & 9.67 \scriptsize{(6.37)}\\
\textbf{ARX} & 47.56 \scriptsize{(3.57)} & 25.11 \scriptsize{(5.33)} & 8.00 \scriptsize{(6.51)}\\

\midrule
\textbf{SVR} & 46.96 \scriptsize{(2.67)} & \underline{\textbf{23.45 \scriptsize{(4.96)}}} & 24.33 \scriptsize{(19.46)}\\
\textbf{GP} & \underline{\textbf{46.24 \scriptsize{(2.86)}}} & 24.65 \scriptsize{(5.69)} & 20.00 \scriptsize{(14.57)}\\
\midrule
\textbf{ELM} & 46.64 \scriptsize{(3.38)} & 24.28 \scriptsize{(5.16)} & 31.50 \scriptsize{(23.92)}\\
\textbf{FFNN} & 46.86 \scriptsize{(2.83)} & 23.82 \scriptsize{(5.39)} & 30.67 \scriptsize{(18.89)}\\
\textbf{LSTM} & 47.56 \scriptsize{(3.05)} & 24.69 \scriptsize{(5.34)} & 21.00 \scriptsize{(14.12)}\\

\bottomrule
\end{tabularx}
}
\end{table*}
\begin{table*}
\caption{Clinical acceptability of glucose-predictive models, averaged over the T1DMS population (with standard deivation), and grouped by prediction horizon (30, 60, and 120 minutes).}
\label{table:cgega_t1dms}
\begin{tabularx}{\textwidth}{@{}l||*{3}{C}|*{3}{C}|*{3}{C}@{}}
\toprule

\multirow{3}{*}{\textbf{Models}} &    \multicolumn{9}{c}{\textbf{CG-EGA}} \\

& \multicolumn{3}{c|}{\textit{Hypoglycemia}} & \multicolumn{3}{c|}{\textit{Euglycemia}} & \multicolumn{3}{c}{\textit{Hyperglycemia}} \\ 

& AP & BE & EP & AP & BE & EP & AP & BE & EP \\

\midrule
\midrule
\multicolumn{10}{c}{\textbf{\textit{Prediction Horizon = 30 minutes}}}\\
\midrule

\textbf{Base} & \footnotesize{73.52} \scriptsize{(14.09)} & \footnotesize{2.18} \scriptsize{(3.86)} & \footnotesize{24.30} \scriptsize{(11.79)} & \footnotesize{95.98} \scriptsize{(1.05)} & \footnotesize{3.47} \scriptsize{(0.85)} & \footnotesize{0.56} \scriptsize{(0.31)} & \footnotesize{85.94} \scriptsize{(9.07)} & \footnotesize{10.88} \scriptsize{(6.87)} & \footnotesize{3.18} \scriptsize{(2.38)}\\
\midrule
\textbf{Poly} & \footnotesize{0.11} \scriptsize{(0.33)} & \underline{\textbf{\footnotesize{0.00} \scriptsize{(0.00)}}} & \footnotesize{99.89} \scriptsize{(0.33)} & \underline{\textbf{\footnotesize{99.27} \scriptsize{(0.33)}}} & \underline{\textbf{\footnotesize{0.68} \scriptsize{(0.31)}}} & \underline{\textbf{\footnotesize{0.05} \scriptsize{(0.06)}}} & \footnotesize{79.88} \scriptsize{(7.80)} & \underline{\textbf{\footnotesize{2.11} \scriptsize{(1.35)}}} & \footnotesize{18.02} \scriptsize{(7.57)}\\
\textbf{AR} & 63.96 \scriptsize{(10.62)} & 14.65 \scriptsize{(7.21)} & 21.39 \scriptsize{(9.03)} & 79.79 \scriptsize{(3.41)} & 18.01 \scriptsize{(2.89)} & 2.20 \scriptsize{(0.59)} & 79.66 \scriptsize{(4.35)} & 18.00 \scriptsize{(3.51)} & 2.34 \scriptsize{(1.00)}\\
\textbf{ARX} & 71.51 \scriptsize{(7.51)} & 10.90 \scriptsize{(5.84)} & 17.60 \scriptsize{(6.59)} & 85.86 \scriptsize{(3.90)} & 12.59 \scriptsize{(3.60)} & 1.55 \scriptsize{(0.43)} & 83.15 \scriptsize{(2.05)} & 15.07 \scriptsize{(1.66)} & 1.78 \scriptsize{(0.76)}\\

\midrule
\textbf{SVR} & 83.27 \scriptsize{(7.05)} & 2.81 \scriptsize{(4.15)} & 13.91 \scriptsize{(5.96)} & 97.63 \scriptsize{(0.47)} & 1.95 \scriptsize{(0.34)} & 0.43 \scriptsize{(0.14)} & \underline{\textbf{96.59 \scriptsize{(0.93)}}} & 2.81 \scriptsize{(0.61)} & \underline{\textbf{0.60 \scriptsize{(0.53)}}}\\
\textbf{GP} & 78.95 \scriptsize{(8.34)} & 4.83 \scriptsize{(3.49)} & 16.22 \scriptsize{(7.22)} & 94.52 \scriptsize{(1.05)} & 4.72 \scriptsize{(0.91)} & 0.76 \scriptsize{(0.26)} & 94.53 \scriptsize{(1.34)} & 4.52 \scriptsize{(1.04)} & 0.95 \scriptsize{(0.49)}\\
\midrule
\textbf{ELM} & 77.84 \scriptsize{(12.82)} & 2.24 \scriptsize{(2.50)} & 19.92 \scriptsize{(12.80)} & 95.19 \scriptsize{(0.96)} & 4.21 \scriptsize{(0.85)} & 0.60 \scriptsize{(0.19)} & 89.59 \scriptsize{(1.74)} & 8.54 \scriptsize{(1.37)} & 1.86 \scriptsize{(0.79)}\\
\textbf{FFNN} & 85.66 \scriptsize{(6.71)} & 3.78 \scriptsize{(3.62)} & 10.56 \scriptsize{(3.76)} & 95.02 \scriptsize{(0.72)} & 4.15 \scriptsize{(0.49)} & 0.83 \scriptsize{(0.33)} & 91.57 \scriptsize{(2.23)} & 6.94 \scriptsize{(1.82)} & 1.49 \scriptsize{(0.58)}\\
\textbf{LSTM} & \underline{\textbf{85.89 \scriptsize{(9.37)}}} & 3.84 \scriptsize{(4.18)} & \underline{\textbf{10.27 \scriptsize{(5.63)}}} & 93.77 \scriptsize{(1.12)} & 5.32 \scriptsize{(0.91)} & 0.91 \scriptsize{(0.32)} & 91.90 \scriptsize{(1.69)} & 6.77 \scriptsize{(1.51)} & 1.33 \scriptsize{(0.66)}\\

\midrule
\midrule
\multicolumn{10}{c}{\textbf{\textit{Horizon de prédiction = 60 minutes}}}\\
\midrule

\textbf{Base}& \footnotesize{39.98} \scriptsize{(22.15)} & \footnotesize{1.50} \scriptsize{(2.52)} & \footnotesize{58.51} \scriptsize{(21.89)} & \footnotesize{95.55} \scriptsize{(1.53)} & \footnotesize{3.60} \scriptsize{(1.26)} & \footnotesize{0.85} \scriptsize{(0.41)} & \footnotesize{81.41} \scriptsize{(7.75)} & \footnotesize{11.30} \scriptsize{(5.00)} & \footnotesize{7.29} \scriptsize{(3.54)}\\
\midrule
\textbf{Poly} & \footnotesize{0.11} \scriptsize{(0.33)} & \underline{\textbf{\footnotesize{0.00} \scriptsize{(0.00)}}} & \footnotesize{99.89} \scriptsize{(0.33)} & \underline{\textbf{\footnotesize{99.27} \scriptsize{(0.33)}}} & \underline{\textbf{\footnotesize{0.68} \scriptsize{(0.31)}}} & \underline{\textbf{\footnotesize{0.05} \scriptsize{(0.06)}}} & \footnotesize{79.88} \scriptsize{(7.80)} & \underline{\textbf{\footnotesize{2.11} \scriptsize{(1.35)}}} & \footnotesize{18.02} \scriptsize{(7.57)}\\
\textbf{AR} & 21.38 \scriptsize{(15.05)} & 8.31 \scriptsize{(6.39)} & 70.31 \scriptsize{(20.77)} & 77.38 \scriptsize{(6.28)} & 20.07 \scriptsize{(5.48)} & 2.55 \scriptsize{(0.87)} & 70.20 \scriptsize{(7.40)} & 22.98 \scriptsize{(4.98)} & 6.82 \scriptsize{(2.58)}\\
\textbf{ARX} & 23.17 \scriptsize{(16.21)} & 9.36 \scriptsize{(5.82)} & 67.47 \scriptsize{(20.74)} & 78.63 \scriptsize{(6.64)} & 18.94 \scriptsize{(5.79)} & 2.43 \scriptsize{(0.93)} & 71.93 \scriptsize{(6.17)} & 22.45 \scriptsize{(4.39)} & 5.63 \scriptsize{(1.92)}\\

\midrule
\textbf{SVR} & 44.20 \scriptsize{(22.04)} & 1.42 \scriptsize{(1.56)} & 54.38 \scriptsize{(23.00)} & 96.67 \scriptsize{(0.69)} & 3.13 \scriptsize{(0.61)} & 0.20 \scriptsize{(0.11)} & \underline{\textbf{91.63 \scriptsize{(4.42)}}} & 7.55 \scriptsize{(3.51)} & \underline{\textbf{0.82 \scriptsize{(1.01)}}}\\
\textbf{GP} & 35.66 \scriptsize{(23.20)} & 1.93 \scriptsize{(2.02)} & 62.41 \scriptsize{(24.44)} & 94.27 \scriptsize{(1.91)} & 5.07 \scriptsize{(1.67)} & 0.65 \scriptsize{(0.29)} & 89.59 \scriptsize{(2.67)} & 8.84 \scriptsize{(2.31)} & 1.57 \scriptsize{(0.75)}\\
\midrule
\textbf{ELM} & 43.07 \scriptsize{(21.59)} & 2.06 \scriptsize{(2.64)} & 54.87 \scriptsize{(22.89)} & 94.37 \scriptsize{(2.02)} & 5.20 \scriptsize{(1.89)} & 0.43 \scriptsize{(0.20)} & 88.30 \scriptsize{(4.00)} & 10.25 \scriptsize{(3.04)} & 1.45 \scriptsize{(1.12)}\\
\textbf{FFNN} & \underline{\textbf{49.55 \scriptsize{(18.33)}}} & 2.39 \scriptsize{(1.93)} & \underline{\textbf{48.06 \scriptsize{(19.06)}}} & 92.19 \scriptsize{(1.93)} & 7.07 \scriptsize{(1.71)} & 0.74 \scriptsize{(0.29)} & 83.68 \scriptsize{(3.45)} & 14.13 \scriptsize{(2.91)} & 2.19 \scriptsize{(0.88)}\\
\textbf{LSTM} & 43.31 \scriptsize{(18.99)} & 3.54 \scriptsize{(2.91)} & 53.16 \scriptsize{(20.07)} & 92.21 \scriptsize{(2.66)} & 6.89 \scriptsize{(2.22)} & 0.90 \scriptsize{(0.47)} & 85.34 \scriptsize{(3.25)} & 11.96 \scriptsize{(3.00)} & 2.70 \scriptsize{(0.90)}\\

\midrule
\midrule
\multicolumn{10}{c}{\textbf{\textit{Horizon de prédiction = 120 minutes}}}\\
\midrule

\textbf{Base} & \footnotesize{18.64} \scriptsize{(14.36)} & \footnotesize{0.56} \scriptsize{(0.77)} & \footnotesize{80.80} \scriptsize{(14.83)} & \footnotesize{94.71} \scriptsize{(2.25)} & \footnotesize{4.29} \scriptsize{(1.32)} & \footnotesize{1.00} \scriptsize{(1.03)} & \footnotesize{73.61} \scriptsize{(7.63)} & \footnotesize{12.04} \scriptsize{(4.45)} & \footnotesize{14.35} \scriptsize{(4.79)}\\
\midrule
\textbf{Poly} & \footnotesize{0.11} \scriptsize{(0.33)} & \underline{\textbf{\footnotesize{0.00}} \scriptsize{(0.00)}} & \footnotesize{99.89} \scriptsize{(0.33)} & \underline{\textbf{\footnotesize{99.27} \scriptsize{(0.33)}}} & \underline{\textbf{\footnotesize{0.68} \scriptsize{(0.31)}}} & \underline{\textbf{\footnotesize{0.05} \scriptsize{(0.06)}}} & \footnotesize{79.88} \scriptsize{(7.80)} & \underline{\textbf{\footnotesize{2.11} \scriptsize{(1.35)}}} & \footnotesize{18.02} \scriptsize{(7.57)}\\
\textbf{AR} & 1.92 \scriptsize{(3.19)} & 1.29 \scriptsize{(2.15)} & 96.79 \scriptsize{(5.32)} & 85.52 \scriptsize{(8.74)} & 12.49 \scriptsize{(7.84)} & 1.99 \scriptsize{(1.03)} & 71.38 \scriptsize{(8.94)} & 14.13 \scriptsize{(5.84)} & 14.50 \scriptsize{(4.38)}\\
\textbf{ARX} & 2.43 \scriptsize{(4.08)} & 1.33 \scriptsize{(2.16)} & 96.24 \scriptsize{(6.21)} & 84.65 \scriptsize{(8.52)} & 13.22 \scriptsize{(7.67)} & 2.14 \scriptsize{(0.95)} & 70.72 \scriptsize{(8.26)} & 14.60 \scriptsize{(5.42)} & 14.67 \scriptsize{(4.51)}\\

\midrule
\textbf{SVR} & \underline{\textbf{38.53 \scriptsize{(18.09)}}} & 2.36 \scriptsize{(2.22)} & \underline{\textbf{59.11 \scriptsize{(19.62)}}} & 95.58 \scriptsize{(1.29)} & 4.10 \scriptsize{(1.14)} & 0.33 \scriptsize{(0.19)} & \underline{\textbf{84.03 \scriptsize{(4.93)}}} & 9.49 \scriptsize{(3.29)} & \underline{\textbf{6.49 \scriptsize{(2.61)}}}\\
\textbf{GP} & 17.53 \scriptsize{(14.60)} & 1.27 \scriptsize{(1.07)} & 81.20 \scriptsize{(15.30)} & 93.34 \scriptsize{(2.90)} & 6.02 \scriptsize{(2.45)} & 0.64 \scriptsize{(0.46)} & 82.53 \scriptsize{(5.03)} & 10.67 \scriptsize{(2.44)} & 6.79 \scriptsize{(3.24)}\\
\midrule
\textbf{ELM} & \footnotesize{17.36} \scriptsize{(13.11)} & \footnotesize{2.69} \scriptsize{(2.70)} & \footnotesize{79.95} \scriptsize{(14.18)} & \footnotesize{89.63} \scriptsize{(4.38)} & \footnotesize{9.16} \scriptsize{(3.68)} & \footnotesize{1.21} \scriptsize{(0.75)} & \footnotesize{76.49} \scriptsize{(7.35)} & \footnotesize{15.72} \scriptsize{(4.22)} & \footnotesize{7.79} \scriptsize{(3.21)}\\
\textbf{FFNN} & 32.83 \scriptsize{(17.69)} & 1.80 \scriptsize{(1.74)} & 65.37 \scriptsize{(18.57)} & 90.90 \scriptsize{(2.44)} & 8.38 \scriptsize{(1.99)} & 0.72 \scriptsize{(0.47)} & 80.15 \scriptsize{(3.70)} & 15.66 \scriptsize{(2.22)} & 4.19 \scriptsize{(1.80)}\\
\textbf{LSTM} & 10.98 \scriptsize{(10.94)} & 2.96 \scriptsize{(2.49)} & 86.06 \scriptsize{(12.68)} & 89.56 \scriptsize{(4.90)} & 9.30 \scriptsize{(4.30)} & 1.14 \scriptsize{(0.65)} & 77.19 \scriptsize{(5.20)} & 14.19 \scriptsize{(2.25)} & 8.63 \scriptsize{(3.33)}\\

\bottomrule

\end{tabularx}

\begin{flushright}
AP: Accurate Prediction; BE: Benign Error; EP: Erroneous Prediction
\end{flushright}

\end{table*}

\begin{table*}
\caption{Clinical acceptability of glucose-predictive models, averaged over the OhioT1DM population (with standard deivation), and grouped by prediction horizon (30, 60, and 120 minutes).}
\label{table:cgega_ohio}
\begin{tabularx}{\textwidth}{@{}l||*{3}{C}|*{3}{C}|*{3}{C}@{}}
\toprule

\multirow{3}{*}{\textbf{Models}} &    \multicolumn{9}{c}{\textbf{CG-EGA}} \\

& \multicolumn{3}{c|}{\textit{Hypoglycemia}} & \multicolumn{3}{c|}{\textit{Euglycemia}} & \multicolumn{3}{c}{\textit{Hyperglycemia}} \\ 

& AP & BE & EP & AP & BE & EP & AP & BE & EP \\

\midrule
\midrule
\multicolumn{10}{c}{\textbf{\textit{Prediction Horizon = 30 minutes}}}\\
\midrule

\textbf{Base} & \footnotesize{39.24} \scriptsize{(16.93)} & \footnotesize{2.82} \scriptsize{(4.17)} & \footnotesize{57.94} \scriptsize{(18.85)} & \footnotesize{90.25} \scriptsize{(3.40)} & \footnotesize{7.11} \scriptsize{(2.44)} & \footnotesize{2.64} \scriptsize{(1.15)} & \underline{\textbf{\footnotesize{84.40} \scriptsize{(3.88)}}} & \footnotesize{11.44} \scriptsize{(2.59)} & \underline{\textbf{\footnotesize{4.16} \scriptsize{(1.83)}}}\\
\midrule
\textbf{Poly} & \footnotesize{0.00} \scriptsize{(0.00)} & \underline{\textbf{\footnotesize{0.00} \scriptsize{(0.00)}}} & \footnotesize{100.00} \scriptsize{(0.00)} & \underline{\textbf{\footnotesize{94.54} \scriptsize{(1.74)}}} & \underline{\textbf{\footnotesize{5.20} \scriptsize{(1.84)}}} & \underline{\textbf{\footnotesize{0.27} \scriptsize{(0.55)}}} & \footnotesize{75.71} \scriptsize{(6.30)} & \underline{\textbf{\footnotesize{7.00} \scriptsize{(2.82)}}} & \footnotesize{17.29} \scriptsize{(5.72)}\\
\textbf{AR} & 38.11 \scriptsize{(21.40)} & 5.30 \scriptsize{(3.87)} & 56.59 \scriptsize{(22.30)} & 85.42 \scriptsize{(5.40)} & 11.47 \scriptsize{(4.22)} & 3.10 \scriptsize{(1.32)} & 79.18 \scriptsize{(2.98)} & 16.06 \scriptsize{(3.16)} & 4.75 \scriptsize{(1.67)}\\
\textbf{ARX} & 38.32 \scriptsize{(23.33)} & 4.88 \scriptsize{(3.92)} & 56.80 \scriptsize{(23.69)} & 85.10 \scriptsize{(5.41)} & 11.67 \scriptsize{(4.25)} & 3.23 \scriptsize{(1.34)} & 78.96 \scriptsize{(2.91)} & 16.26 \scriptsize{(3.00)} & 4.78 \scriptsize{(1.69)}\\

\midrule
\textbf{SVR} & 46.89 \scriptsize{(23.72)} & 6.62 \scriptsize{(4.97)} & 46.49 \scriptsize{(23.87)} & 86.44 \scriptsize{(4.25)} & 10.64 \scriptsize{(3.22)} & 2.92 \scriptsize{(1.25)} & 80.90 \scriptsize{(3.31)} & 14.64 \scriptsize{(3.03)} & 4.46 \scriptsize{(1.90)}\\
\textbf{GP} & 46.00 \scriptsize{(26.35)} & 6.31 \scriptsize{(3.93)} & 47.69 \scriptsize{(27.78)} & 84.61 \scriptsize{(5.39)} & 12.22 \scriptsize{(4.16)} & 3.18 \scriptsize{(1.41)} & 78.35 \scriptsize{(3.63)} & 16.83 \scriptsize{(3.28)} & 4.82 \scriptsize{(1.60)}\\
\midrule
\textbf{ELM} & 34.81 \scriptsize{(23.43)} & 6.81 \scriptsize{(4.22)} & 58.39 \scriptsize{(23.88)} & 78.85 \scriptsize{(4.32)} & 17.25 \scriptsize{(3.18)} & 3.91 \scriptsize{(1.57)} & 73.32 \scriptsize{(4.41)} & 20.79 \scriptsize{(3.54)} & 5.89 \scriptsize{(1.73)}\\
\textbf{FFNN} & \underline{\textbf{51.88 \scriptsize{(21.65)}}} & 3.58 \scriptsize{(3.23)} & \underline{\textbf{44.54 \scriptsize{(21.63)}}} & 82.57 \scriptsize{(5.22)} & 13.73 \scriptsize{(4.00)} & 3.70 \scriptsize{(1.40)} & 74.60 \scriptsize{(4.19)} & 19.55 \scriptsize{(3.64)} & 5.84 \scriptsize{(2.28)}\\
\textbf{LSTM} & 38.37 \scriptsize{(23.17)} & 3.97 \scriptsize{(3.72)} & 57.67 \scriptsize{(24.23)} & 83.78 \scriptsize{(5.33)} & 12.70 \scriptsize{(4.06)} & 3.52 \scriptsize{(1.47)} & 76.86 \scriptsize{(3.70)} & 17.87 \scriptsize{(2.73)} & 5.27 \scriptsize{(2.21)}\\

\midrule
\midrule
\multicolumn{10}{c}{\textbf{\textit{Horizon de prédiction = 60 minutes}}}\\
\midrule

\textbf{Base} & \underline{\textbf{\footnotesize{20.00} \scriptsize{(15.92)}}} & \footnotesize{2.86} \scriptsize{(3.12)} & \underline{\textbf{\footnotesize{77.14} \scriptsize{(15.32)}}} & \footnotesize{88.27} \scriptsize{(4.23)} & \footnotesize{8.22} \scriptsize{(2.57)} & \footnotesize{3.51} \scriptsize{(1.72)} & \underline{\textbf{\footnotesize{81.77} \scriptsize{(4.65)}}} & \footnotesize{12.06} \scriptsize{(2.43)} & \underline{\textbf{\footnotesize{6.17} \scriptsize{(2.33)}}}\\
\midrule
\textbf{Poly} & \footnotesize{0.00} \scriptsize{(0.00)} & \underline{\textbf{\footnotesize{0.00} \scriptsize{(0.00)}}} & \footnotesize{100.00} \scriptsize{(0.00)} & \underline{\textbf{\footnotesize{94.77} \scriptsize{(1.85)}}} & \underline{\textbf{\footnotesize{5.13} \scriptsize{(1.90)}}} & \underline{\textbf{\footnotesize{0.10} \scriptsize{(0.18)}}} & \footnotesize{75.87} \scriptsize{(6.23)} & \underline{\textbf{\footnotesize{6.99} \scriptsize{(2.86)}}} & \footnotesize{17.14} \scriptsize{(5.74)}\\
\textbf{AR} & 7.58 \scriptsize{(7.38)} & 1.67 \scriptsize{(2.09)} & 90.74 \scriptsize{(8.74)} & 83.24 \scriptsize{(5.01)} & 12.71 \scriptsize{(3.84)} & 4.05 \scriptsize{(1.34)} & 74.91 \scriptsize{(4.36)} & 17.71 \scriptsize{(2.81)} & 7.39 \scriptsize{(3.07)}\\
\textbf{ARX} & 7.67 \scriptsize{(7.42)} & 1.97 \scriptsize{(2.30)} & 90.36 \scriptsize{(8.90)} & 82.12 \scriptsize{(4.92)} & 13.69 \scriptsize{(3.81)} & 4.19 \scriptsize{(1.28)} & 73.78 \scriptsize{(4.64)} & 18.54 \scriptsize{(3.10)} & 7.67 \scriptsize{(3.09)}\\

\midrule
\textbf{SVR} & 16.92 \scriptsize{(16.48)} & 3.07 \scriptsize{(2.56)} & 80.02 \scriptsize{(17.29)} & 82.96 \scriptsize{(4.09)} & 13.24 \scriptsize{(3.14)} & 3.80 \scriptsize{(1.17)} & 75.68 \scriptsize{(4.87)} & 17.41 \scriptsize{(3.17)} & 6.91 \scriptsize{(2.51)}\\
\textbf{GP} & 12.99 \scriptsize{(9.78)} & 2.27 \scriptsize{(2.82)} & 84.75 \scriptsize{(12.09)} & 82.07 \scriptsize{(5.37)} & 13.85 \scriptsize{(4.13)} & 4.09 \scriptsize{(1.43)} & 74.26 \scriptsize{(5.21)} & 18.61 \scriptsize{(3.06)} & 7.13 \scriptsize{(2.81)}\\
\midrule
\textbf{ELM} & 8.75 \scriptsize{(7.99)} & 3.48 \scriptsize{(3.57)} & 87.77 \scriptsize{(11.16)} & 77.24 \scriptsize{(3.77)} & 18.37 \scriptsize{(2.89)} & 4.40 \scriptsize{(1.41)} & 71.96 \scriptsize{(5.30)} & 20.16 \scriptsize{(3.02)} & 7.87 \scriptsize{(2.93)}\\
\textbf{FFNN} & 10.93 \scriptsize{(10.14)} & 1.04 \scriptsize{(1.32)} & 88.04 \scriptsize{(11.39)} & 79.15 \scriptsize{(5.31)} & 16.35 \scriptsize{(4.12)} & 4.50 \scriptsize{(1.35)} & 69.43 \scriptsize{(6.23)} & 22.33 \scriptsize{(3.53)} & 8.25 \scriptsize{(3.52)}\\
\textbf{LSTM} & 3.48 \scriptsize{(5.14)} & 1.07 \scriptsize{(1.09)} & 95.45 \scriptsize{(5.83)} & 82.40 \scriptsize{(5.19)} & 13.59 \scriptsize{(3.89)} & 4.01 \scriptsize{(1.46)} & 72.33 \scriptsize{(6.42)} & 19.43 \scriptsize{(3.19)} & 8.24 \scriptsize{(3.78)}\\

\midrule
\midrule
\multicolumn{10}{c}{\textbf{\textit{Horizon de prédiction = 120 minutes}}}\\
\midrule

\textbf{Base} & \footnotesize{7.76} \scriptsize{(6.78)} & \footnotesize{0.91} \scriptsize{(1.02)} & \footnotesize{91.33} \scriptsize{(6.98)} & \footnotesize{85.06} \scriptsize{(3.88)} & \footnotesize{9.09} \scriptsize{(2.92)} & \footnotesize{5.85} \scriptsize{(1.08)} & \underline{\textbf{\footnotesize{78.58} \scriptsize{(5.01)}}} & \footnotesize{12.07} \scriptsize{(2.41)} & \underline{\textbf{\footnotesize{9.36} \scriptsize{(3.25)}}}\\
\midrule
\textbf{Poly} & \footnotesize{0.00} \scriptsize{(0.00)} & \underline{\textbf{\footnotesize{0.00} \scriptsize{(0.00)}}} & \footnotesize{100.00} \scriptsize{(0.00)} & \underline{\textbf{\footnotesize{94.80} \scriptsize{(1.83)}}} & \underline{\textbf{\footnotesize{5.10}} \scriptsize{(1.87)}} & \underline{\textbf{\footnotesize{0.10} \scriptsize{(0.15)}}} & \footnotesize{75.88} \scriptsize{(6.16)} & \underline{\textbf{\footnotesize{7.00} \scriptsize{(2.89)}}} & \footnotesize{17.12} \scriptsize{(5.63)}\\
\textbf{AR} & 0.00 \scriptsize{(0.00)} & 0.00 \scriptsize{(0.00)} & 100.00 \scriptsize{(0.00)} & 85.55 \scriptsize{(3.22)} & 11.18 \scriptsize{(2.57)} & 3.27 \scriptsize{(0.99)} & 74.94 \scriptsize{(5.61)} & 14.04 \scriptsize{(1.86)} & 11.02 \scriptsize{(5.12)}\\
\textbf{ARX} & 0.55 \scriptsize{(1.23)} & 0.22 \scriptsize{(0.48)} & 99.23 \scriptsize{(1.72)} & 86.09 \scriptsize{(3.54)} & 10.85 \scriptsize{(2.88)} & 3.06 \scriptsize{(0.85)} & 74.63 \scriptsize{(6.55)} & 13.28 \scriptsize{(2.37)} & 12.09 \scriptsize{(5.02)}\\

\midrule
\textbf{SVR} & \underline{\textbf{10.71 \scriptsize{(10.54)}}} & 2.75 \scriptsize{(4.48)} & \underline{\textbf{86.53 \scriptsize{(14.62)}}} & 84.10 \scriptsize{(2.66)} & 12.53 \scriptsize{(2.92)} & 3.36 \scriptsize{(0.88)} & 75.52 \scriptsize{(4.36)} & 14.06 \scriptsize{(2.22)} & 10.42 \scriptsize{(3.55)}\\
\textbf{GP} & 2.28 \scriptsize{(3.50)} & 0.43 \scriptsize{(0.97)} & 97.29 \scriptsize{(4.38)} & 83.59 \scriptsize{(3.37)} & 12.87 \scriptsize{(2.69)} & 3.54 \scriptsize{(1.23)} & 74.82 \scriptsize{(5.54)} & 14.83 \scriptsize{(2.17)} & 10.35 \scriptsize{(4.01)}\\
\midrule
\textbf{ELM} & 0.46 \scriptsize{(0.59)} & 0.05 \scriptsize{(0.10)} & 99.49 \scriptsize{(0.68)} & 79.85 \scriptsize{(2.62)} & 16.40 \scriptsize{(1.51)} & 3.75 \scriptsize{(1.22)} & 72.49 \scriptsize{(5.89)} & 15.92 \scriptsize{(1.68)} & 11.59 \scriptsize{(4.57)}\\
\textbf{FFNN} & 3.55 \scriptsize{(4.69)} & 1.18 \scriptsize{(1.54)} & 95.27 \scriptsize{(5.54)} & 78.04 \scriptsize{(3.26)} & 18.03 \scriptsize{(2.48)} & 3.93 \scriptsize{(1.06)} & 69.79 \scriptsize{(6.23)} & 19.11 \scriptsize{(2.94)} & 11.10 \scriptsize{(4.32)}\\
\textbf{LSTM} & 1.06 \scriptsize{(1.99)} & 0.91 \scriptsize{(2.03)} & 98.04 \scriptsize{(4.01)} & 82.39 \scriptsize{(3.43)} & 13.92 \scriptsize{(2.48)} & 3.68 \scriptsize{(1.34)} & 72.89 \scriptsize{(5.68)} & 16.29 \scriptsize{(2.20)} & 10.82 \scriptsize{(4.41)}\\

\bottomrule

\end{tabularx}

\begin{flushright}
AP: Accurate Prediction; BE: Benign Error; EP: Erroneous Prediction
\end{flushright}

\end{table*}

First, only the Poly model is performing worse than our baseline (having a worse accuracy for both datasets and all prediction horizons). While this shows that the day-to-day variability is too important for a simple model based on time to work, this also shows that the simulated data are irregular enough, as a result of the randomization of the meal and insulin quantities and timings. However, the Poly model displays the best TG scores and the average dynamics are well captured (CG-EGA in euglycemia and hyperglycemia) since they are more or less the same day after day. Nonetheless, the model is still not usable since it is unable to detect hypoglycemia with over 99\% of EP for every PH due to its low accuracy. Besides, as for the TG scores, this shows that we need to be careful when using the TG metric, as it cannot be used on its own to assess the forecasting ability of a model but should always be used in combination with other standard accuracy metrics. 

The results displayed by the AR and ARX models are quite similar, with the ARX being slightly better (this shows the interest of using additional information, such as past insulin boluses or CHO intakes, to predict future glucose values). However, both models are significantly outclassed by the other remaining models. They are less accurate (higher RMSE and MAPE), provide less anticipation (lower TG), and are overall less safe (CG-EGA). This is due to their inherent simplicity (linear regression) hindering the modeling of glucose variations.

When it comes to the SVR and GP models, we can witness a lot of improvement when compared to the linear regression models (Poly, AR, ARX) for every PH and for every metric. The clinical acceptability of the GP model deteriorates more rapidly than the SVR model, making the use of the SVR model more interesting as it has overall very good performances. This shows the usefulness of the non-linear transformation of the input space through the use of kernels, transformation the SVR model seems to take advantage of really well.

% As for the neural-network-based models, they are on par with the kernel-based solutions for all the prediction horizons, and even outclassing them at long-term forecasting. Moreover, they have the best TG scores accross all the models. Between the ELM, the FFNN, and the LSTM models, the LSTM model seems to be the best at short-term predictions, while the FFNN and ELM are especially accurate at longer PH. Regarding the clinical acceptability, the FFNN model seems to be well suited for hypoglycemia detection but at the same time seems to produce a higher benign error rate in the euglycemia and hyperglycemia regions. On the other hand, the LSTM model displays the opposite behavior, being unsuccessful in the hypoglycemia region, and very successfull in the other ones. 

As for the neural-network-based models, apart from the ELM whose results are poor overall, their results fall between these of the autoregressive the kernel-based models. The FFNN and the LSTM models both shine in different and complementary areas. While the LSTM is good at making short-term predictions and predictions in hyperglycemia, the FFNN performs best at longer PH and in hypoglycemia. This difference in performance could be explained by their inherent nature. While the LSTM model makes predictions which are coherent between each other's (because of its ability to remember past events), this ability might not be helpful in all the situations \cite{gers2002applying}. In our context, remembering CHO intakes and insulin boluses helps the LSTM model to be one of the best models in the hyperglycemia regions. The LSTM model seems to correctly capture the global dynamics of glucose through its variations, but has a harder time predicting rare events such as hypoglycemia (compared to the FFNN model).

\begin{figure*}
    \centering
    \includegraphics[width=\textwidth]{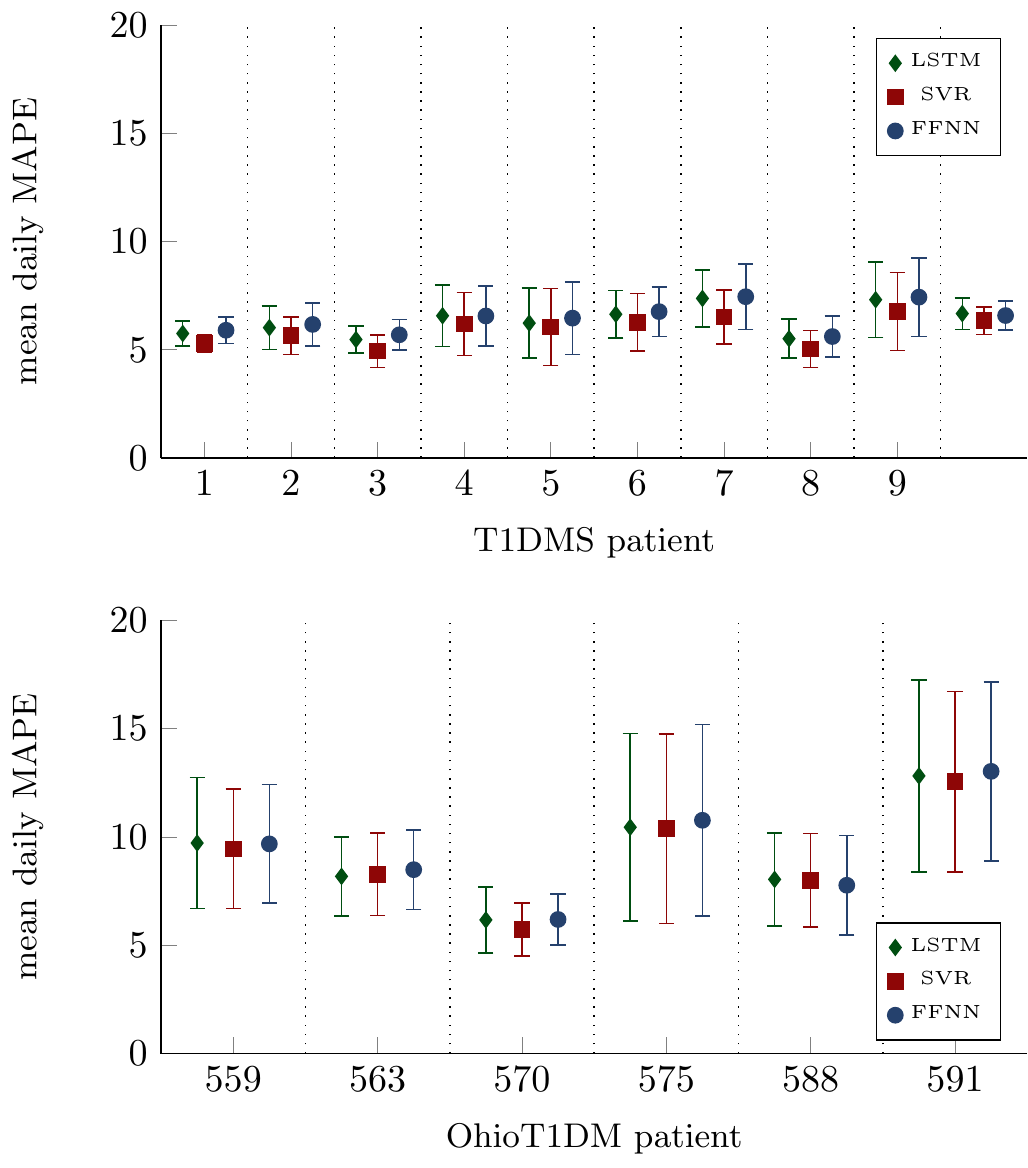}
    \caption{Per patient (ID) mean and standard deviation of daily MAPE (PH of 30 minutes) for the T1DMS dataset (top) and the OhioT1DM dataset (bottom)}
    \label{fig:intrainter_both}
\end{figure*}

With the help of Figure \ref{fig:intrainter_both}, we provide more insight on the intra-/inter-patient variability. For both datasets, some patients are easier to make predictions for, and some patients have higher daily variability than others. However, while we show that the day-to-day variability of T1DMS patients is enough to prevent a simple time model to perform well, the graph shows that the real OhioT1DM patients seems to have an even higher day-to-day variability. This difference in variability can be explained by several factors. First, real glucose signals are not only impacted by insulin infusions and CHO intakes (as the synthetic glucose signals are) but also by other factors such as physical activity, sleep, or emotions. Second, there are a lot of missing data in the OhioT1DM dataset, making the predictions for certain days very hard to make. That being said, because the relative strengths of the models are the same for both datasets, we can conclude that the evaluation of a model forecasting ability can be done on T1DMS even though its data are synthetic and thus not fully representative of real data.

\section{Conclusion}

Data availability is a big challenge that holds the glucose prediction research field back. T1DMS, the diabetic metabolic simulator made by the universities of Virginia and Padova, alongside the recently released OhioT1DM dataset present a unique opportunity to gather researchers around the same data to work on improving the forecast of future glucose values.

In this paper, we presented a methodology to benchmark the machine-learning-based glucose-predictive models. After going through the exhaustive details on how to run the benchmark from the acquisition of data to the evaluation of the models, we present the results of nine models taken from the glucose-prediction literature.

The results showed that, while more in favor of complex, non-linear, predictive models (kernel-based or neural-network-based regression), no model outperforms the others in every category. Every model has its own strengths and weaknesses. The support vector regression model is the most accurate model when the prediction horizon is short-term. Finally, as the relative performances of the models are the same for the virtual (T1DMS) and real (OhioT1DM) datasets, we conclude that the virtual data obtained through T1DMS and our scenario can be used to evaluate the ability of the models to forecast future glucose values.

However, these results also show the difficulty of predicting future glucose values, in particular at high prediction horizons, and the need of the GLYFE benchmark to incentivize comparisons between studies. We hope that our work will serve as a basis for future work in the glucose prediction field.

\section*{Acknowledgment}

This work is supported by the ``IDI 2017'' project funded by the IDEX Paris-Saclay, ANR-11-IDEX-0003-02.

\bibliography{IEEEabrv.bib,bibtex.bib}
\bibliographystyle{IEEEtran}

% \bibliography{bibtex.bib}
% \bibliographystyle{IEEEtran}

\ifCLASSOPTIONcaptionsoff
  \newpage
\fi

\end{document}